\documentclass[a4paper,12pt]{article}
\usepackage[english]{babel}
\usepackage[utf8]{inputenc}
\usepackage[T1]{fontenc}
\usepackage{graphics}
\usepackage{booktabs}
\usepackage{amsmath}
\usepackage{inputenc}
\usepackage{amsthm}
\usepackage{amssymb}
\usepackage{setspace}
\usepackage{xcolor}
\usepackage{array}
\usepackage{tabularx}
\usepackage{makecell}
\usepackage[percent]{overpic}
\usepackage{natbib}
\usepackage{graphicx}
\usepackage{longtable}
\usepackage{mathtools}
\usepackage{subfig}
\usepackage{bbm}
\usepackage{rotating}
\usepackage{multirow}
\usepackage{url}
\usepackage{varioref}
\usepackage[blocks]{authblk}
\usepackage[font=footnotesize,format=hang]{caption}
\usepackage[a4paper,top=3cm,bottom=3cm,left=2.5cm,right=2.5cm,%
bindingoffset=5mm]{geometry}

\theoremstyle{remark}

\theoremstyle{plain}

\usepackage{timet}
\usepackage{amsmath}
\usepackage{lineno}
\usepackage{hyperref}
\usepackage{graphicx}
\usepackage{xcolor}

\usepackage{lipsum}
\usepackage{mathtools, cuted}
\usepackage{lineno}
\usepackage{setspace} 
\usepackage{rotating}
\usepackage{amsmath}

\usepackage{soul}

\usepackage{xr}

\begin{document}

\title{\textbf{Reconciling the irreconcilable: window-based versus stochastic declustering algorithms}}
\author[Spassiani \emph{et al.}]
  {I. Spassiani$^1$\thanks{Corresponding author}, S. Gentili$^2$, R. Console$^{3,1}$, M. Murru$^1$, M. Taroni$^1$, G. Falcone$^1$\\
  $^1$ Istituto Nazionale di Geofisica e Vulcanologia (INGV), Via di Vigna Murata 605, 00143 Rome, Italy\\
  $^2$ National Institute of Oceanography and Applied Geophysics - OGS, Udine, Italy\\
  $^3$ Center of Integrated Geomorphology for the Mediterranean Area, Potenza, Italy
  }
\date{}

\maketitle

\begin{abstract}
Short-term earthquake clustering is one of the most important features of seismicity. Clusters are identified using various techniques, generally deterministic and based on spatio-temporal windowing. Conversely, the leading rail in short-term earthquake forecasting has a probabilistic view of clustering, usually based on the Epidemic Type Aftershock Sequence (ETAS) models. In this study we compare seismic clusters, identified by two different deterministic window-based techniques, with the ETAS probabilities associated with any event in the clusters, thus investigating the consistency between deterministic and probabilistic approaches. The comparison is performed by considering, for each event in an identified cluster, the corresponding probability of being independent and the expected number of triggered events according to ETAS. 
Results show no substantial differences between the cluster identification procedures, and an overall consistency between the identified clusters and the relative events' ETAS probabilities. 
\end{abstract}

\section*{Keywords}
Window-based clustering techniques, stochastic declustering, statistical methods, earthquake dynamics, cluster detection.

\section{Introduction}
The Epidemic Type Aftershock Sequence (ETAS) model represents a benchmark in statistical seismology, very often used to forecast earthquake sequences at various spatio-temporal scales \citep{ogata:1998,console:2001,console:2003,console:2007,lombardi:2010,omi:2014}. It is a branching, self-exciting, Hawkes process, according to which any seismic event may generate its own aftershocks independently of the other events and of the background (``spontaneous'', independent) ones. The ETAS model is based on three simple constitutive laws, i.e. (i) the Omori-Utsu for the aftershocks' temporal decay, (ii) a spatial distribution usually of Gaussian type, and (iii) the exponential Gutenberg-Richter law for the events’ frequency magnitudes \citep{omori:1895,gutenberg:1944,utsu:1957,ogata:1988,ogata:1998,zhuang:2002}. This model specifically provides the occurrence probability of an earthquake in a fixed space-time-magnitude domain.

One of the key strengths of ETAS relies in its capability to account for the main characteristic of seismicity, that is, the events' clustering in space and time. Due to its probabilistic nature, the ETAS model is very often used to decluster an earthquake catalog through a stochastic approach. However, it is important to stress that a stochastic declustered catalog is not unique, as it depends on the random numbers used to identify the events that form the background component of seismicity \citep{zhuang:2002}. Indeed, precisely because of its probabilistic nature, selecting a specific probability threshold to identify clusters by means of ETAS leads to a distortion of the hypothesis upon which this model is built. Instead, the cluster identification procedures typically adopted in the literature rest on deterministic window-based methods, according to which some constitutive equations are selected to set up the spatio-temporal extent of any cluster. Several different window-based methods have been proposed in the literature, mainly differing on the specific set of equations adopted (e.g. \cite{gardner:1974,keilis:1980,reasenberg:1985,vanstiphout:2012}). 

The main goal of this paper is to compare seismic clusters identified by two different deterministic window-based techniques, to the probabilities that the ETAS model associates with any event in the clusters. A few recent studies in the literature already investigated some classification similarities and differences between the Nearest‐Neighbor \citep{zaliapin:2013} and the Stochastic \citep{zhuang:2002} earthquake declustering methods, mainly by using spatio-temporal statistical measures and tools from network analysis \citep{varini:2020,benali:2023}. Instead, in this paper we develop an automatic approach that, based on two simple checks, is able to assess the consistency of the identified clusters with the ETAS rates. It is worth mentioning that, if, on the one hand, window-based models create clusters starting from a strong event, usually named ``mainshock'', on the other hand, in the ETAS model events’ are not labeled as ``mainshocks, aftershocks or foreshocks''. ETAS just distinguishes between background (independent) and triggered seismicity. Also in light of this difference, it is interesting to see if and how window-based clusters comply with the ETAS probabilities. The specific formulation of the ETAS rate density of earthquakes we consider in this paper is given in the Appendix~\ref{sec:appa} (see also \cite{console:2010}). 

The clusters we consider to assess consistency with the ETAS probabilities  are identified by applying the cluster identification module of NESTOREv1.0 software \citep{gentili:2023}, which allows us to make a window-based cluster identification by specifying the desired laws. In particular, we consider here two different sets of laws for determining clusters: the ones by \cite{gardner:1974}, mostly used in the literature, and the ones by \cite{uhrhammer:1986} and \cite{lolli:2003}, as in \cite{gentili:2017}.

\section{The earthquake catalog}
The earthquake catalog to which we apply the clustering procedures is ISIDe (Italian Seismological Instrumental and Parametric Data-Base, \url{http://terremoti.ingv.it/ISIDe}) from 2005 April 18, to 2021 April 30, over the entire Italian territory and some neighboring areas covered by the Italian seismic network (see Figure~\ref{fig:figureS1} in the Supplement). The minimum and maximum magnitudes in the catalog are ML 0.9 and ML 6.1, respectively. The completeness threshold, estimated  for this catalog by \cite{Zhuang:2019}, is  ML 2.9, above which we count a total of 5084 events. For simplicity, in what follows we will name the events in the ISIDe catalog as ``I-events''.

In order to assess clusters' consistency with the ETAS model (e.g., \cite{console:2001,console:2003}), we need to associate to any I-event both the corresponding ETAS probability of being independent, and the number of triggered events expected by the ETAS model. The first quantity is obtained as the ratio between the background event’s rate and the total rate; the second quantity is instead simply the expected value of aftershocks per each event. To obtain these quantities, we use the algorithms developed by \cite{console:2010} with the parameters estimated by means of the Maximum Likelihood Estimation (MLE) technique, performed over the entire catalog. The specific parameter values found in the last iteration are $(f_r, K, d_0, q, c, p, \alpha)= (0.25, 0.12, 1.18, 1.89, 5.7$ E-$03, 1.09, 0.54)$. The choice of not considering space-time varying parameters is to be coherent with the window-based methods, for which we use fixed equations for the entire territory. Besides, this choice is not expected to influence our results because we are not providing a forecast of seismicity, but performing a retrospective analysis.

\section{Cluster identification methods}
Most window-based cluster identification methods have a similar algorithm. They start from an equation defining the space and time triggering area for the mainshock, set a minimum threshold for the mainshock magnitude and define the cluster of a mainshock, $m_i$, as ``the set of all the earthquakes after $m_i$ within its triggering area''. If the cluster of a given mainshock contains a larger earthquake $m_j$, the clusters of $m_i$ and $m_j$ are merged, $m_j$ becomes the cluster's mainshock, and the events before $m_j$ become foreshocks. The space window is a circular area around the mainshock, and the radius of the circle depends on the cluster identification method: different equations may be applied depending also on the seismotectonics of the region; both the radius and the time window are generally a function of the mainshock magnitude \citep{vanstiphout:2012}.   

\subsection{The two approaches used to identify clusters}
As anticipated in the Introduction, in this paper we perform cluster identification by using a module implemented in the package NESTOREv1.0 \citep{gentili:2023}. This package allows us to detect clusters of seismicity by choosing the equations which define the space and time triggering area. In addition, it selects as foreshocks all the events before the mainshock within a radius arbitrarily set to 1.5 the radius of the mainshock, and a time window of 1 month. However, here we will not take into account the foreshocks, to avoid possible multiple assignments of the same events to clusters close in space and time.  For the sake of clarity, we stress that a cluster must obviously contain at least two events. The two specific sets of equations we consider here to identify the clusters, explicitly given in the Appendix~\ref{sec:appb}, are those by \cite{uhrhammer:1986}-\cite{lolli:2003} and \cite{gardner:1974}; hereafter, the relative clusters will be obtained by setting a minimum threshold for mainshock magnitude $M_m$ equal to 4.0, and will be named ``ULG-clusters'' and ``GK-clusters'' respectively. The specif choice of $M_m=4.0$ may somehow hinder some results of comparison with ETAS, however it is due to the fact that, for smaller maishocks, the window-based methods more likely fail to assign the ``true'' aftershocks to their mainshock's cluster, assigning instead the events that occur there by chance. This is a direct consequence of the typical seismic activity in Italy, which is for a quite high percentage of background type (i.e. random). In any case, to be sure that our results are not biased by the specific choice of $M_m=4.0$, we repeated the analysis by setting a minimum mainshock magnitude equal to the completeness threshold 2.9. The results we obtained are in their entirety consistent to those illustrated here.

Figure~\ref{fig:figureS2} in the Supplement shows the comparison of the sets of Equations~\ref{eqn:ulg} and~\ref{eqn:gk} (see the Appendix~\ref{sec:appb}) in space and time. The radius of the ULG-clusters is smaller than the GK- ones for $M_m < 6.3$; since the analyzed clusters in Italy have a mainshock with a magnitude lower than this value (we recall that the maximum magnitude in the ISIDe catalog we consider is ML 6.1), narrower clusters are generally expected by considering the ULG method. This effect is only partially compensated by the higher duration of the ULG-clusters for $M_m < 4.8$.

\subsection{Identification of ULG- and GK-clusters, and their comparison with ETAS independence probability}
The clusters identified by implementing the method ULG in NESTOREv1.0 (ULG-clusters) are 79 in total. The number of clusterized events is 2516; since the cardinality of the ISIDe catalog is 5084 $($ML$\, \ge 2.9)$, this means that there are 2568 ``single'' events, not associated with any ULG-cluster. The main information relative to the ULG-clusters with strongest magnitude ML 4.5 (22 in total) is given in Table~\ref{tab:tableS1} of the Supplement, where it can be observed that there are some cases of temporal overlapping. In Figure~\ref{fig:figureS3} of the Supplement we show all the ULG-clusters in different colors (events are represented as circles). 

\begin{sidewaysfigure}
    \centering   
    \includegraphics[trim=0.01cm 0.1cm 0.01cm 0.01cm,clip,width=22cm]{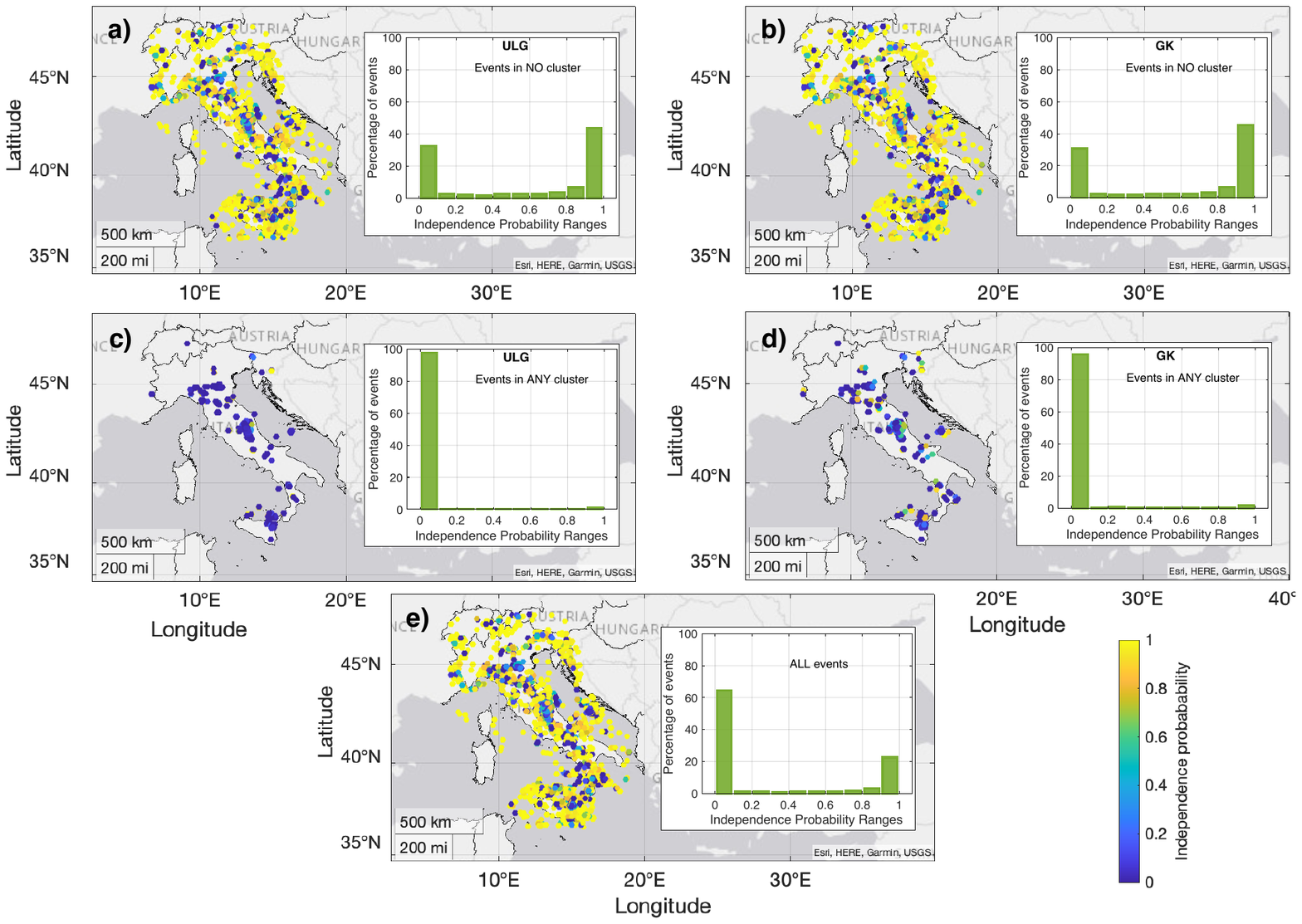}
    \caption{Seismic maps of the ISIDe events coloured according to their independence ETAS probability. Top, middle and bottom panels contain the events that do not belong to any cluster, that belong to some cluster, and all the events together, respectively. The left (right) panels concern the ULG- (GK-) clusters. In the insets, corresponding histograms relative to the ETAS independence probabilities associated with the events.}
    \label{fig:figure1}
\end{sidewaysfigure}

The 2568 I-events that do not belong to any ULG-cluster (hereafter, ``I-not-ULG'') are mapped in panel a) of Figure~\ref{fig:figure1}, where they are coloured according to their independence ETAS probability. Panels c) and e) in the same figure concern instead the 2516 clusterized events and all the 5084 events together, respectively. As expected, clusterized events are much less sparse than the I-not-ULG ones, such that they seem to be much less than these latter and much less than their effective number 2516: this obviously depends just on graphical rendering.

\begin{figure}
    \hspace{-2.1cm}
    \includegraphics[trim=0.01cm 4cm 0.01cm 3cm,clip,width=20cm]{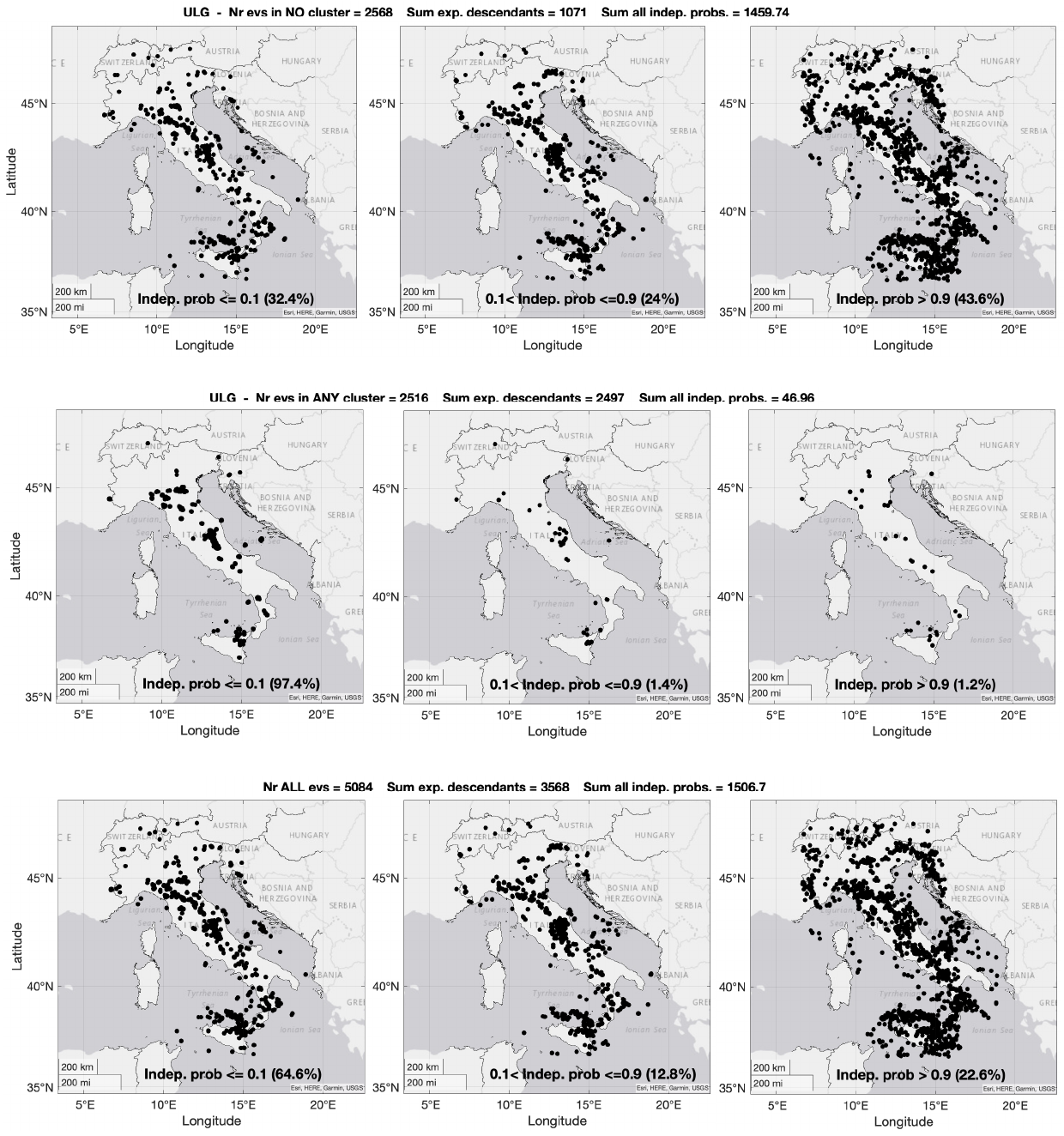}
    \caption{Seismic maps of the ISIDe events that do not belong to any ULG-cluster (top) or belong to any ULG-cluster (middle). The bottom panels refer instead to all the ISIDe events together. Left, middle and right panels for independence probability $\le 0.1$, in $(0.1, 0.9]$, and $>0.9$, respectively.}
    \label{fig:figure2}
\end{figure}

When we consider bins of 0.1 independence probability, we obtain a prevalence of I-not-ULG events in $[0, 0.1]$ and in $(0.9, 1]$, with a slightly higher numerosity in this latter bin, likely representing the foreshocks we have not accounted for to avoid possible multiple assignments (see Subsection above). As expected, the great majority of clusterized events have instead a very low probability of being independent $(< 0.1)$, with the exception of a few of them reasonably representing the mainshocks of the clusters. This is shown in the insets of respectively panels a) and c) of Figure~\ref{fig:figure1}, and confirmed by the maps in top and middle panels of Figure~\ref{fig:figure2}, where we plot separately the events with independence probability $\le 0.1$,  $\le 0.9\; \& > 0.1$, $> 0.9$ (respectively in left, middle and right panels). The relative percentages are also indicated, together with the number of events considered, and the sums of their relative expected descendants and independence ETAS probabilities. The same quantities are reported also in Table~\ref{tab:tableS2} of the Supplement. The inset in panel e) of Figure~\ref{fig:figure1} and the bottom panels in Figure~\ref{fig:figure2} show the same results but relative to all the events in the ISIDe catalog. A focus on the spatio-temporal location of the I-not-ULG events with independence probability in the $[0, 0.1]$ bin is instead given in the top panel of Figure~\ref{fig:figureS4} of the Supplement, where the events are coloured according to the year of occurrence and have size which increases with the magnitude (from ML 2.9 to ML 5.4).  Interestingly, it is not possible to identify a specific characteristic they have (e.g. the majority of them occur in the borders, or have a large magnitude), thus an additional analysis is required to investigate this result. In the top left and middle left panels of Figure~\ref{fig:figureS5} in the Supplement we finally show the frequency magnitude distributions FMDs (both the probability and cumulative densities) of the I-not-ULG and ULG-clusterized events, respectively. The tail of this latter is longer than the former, and this obviously derives from the fact that the stronger events are very likely the clusters' mainshocks. We also stress that the top panel presents a step in correspondence of magnitude 4.0, but this is an artifact due to the minimum threshold for mainshock magnitude we set equal to this value ($M_m$ min $=$ 4.0): clusters with $M_m < 4.0$ do not exist, and those with a just larger mainshock magnitude appear overestimated. When repeating the same analysis with a minimum mainshock magnitude equal to the completeness threshold, this effect is indeed much reduced. Finally, as before, the bottom panel in Figure~\ref{fig:figureS5} of the Supplement concerns instead the FMD of all the I-events together.    

The GK-clusters identified by implementing the NESTOREv1.0 software with Equations~\ref{eqn:gk} (see the Appendix~\ref{sec:appb}) are 82 in total, and are plotted in Figure~\ref{fig:figureS6} of the Supplement. In Table~\ref{tab:tableS3} of this latter file we report the main information relative to the clusters with strongest magnitude ML 4.5. Also in this case, these latter are a total of 22, and some of them do overlap in time. The total number of clusterized events is 2653. Recalling that the cardinality of the ISIDe catalog is 5084, we deduce that there are 2431 ``single'' events not associated with any GK-cluster (i.e. 2431 ``I-not-GK events''). 

\begin{figure}
    \hspace{-2.15cm}
    \includegraphics[trim=0.01cm 12cm 0.01cm 2.2cm,clip,width=20cm]{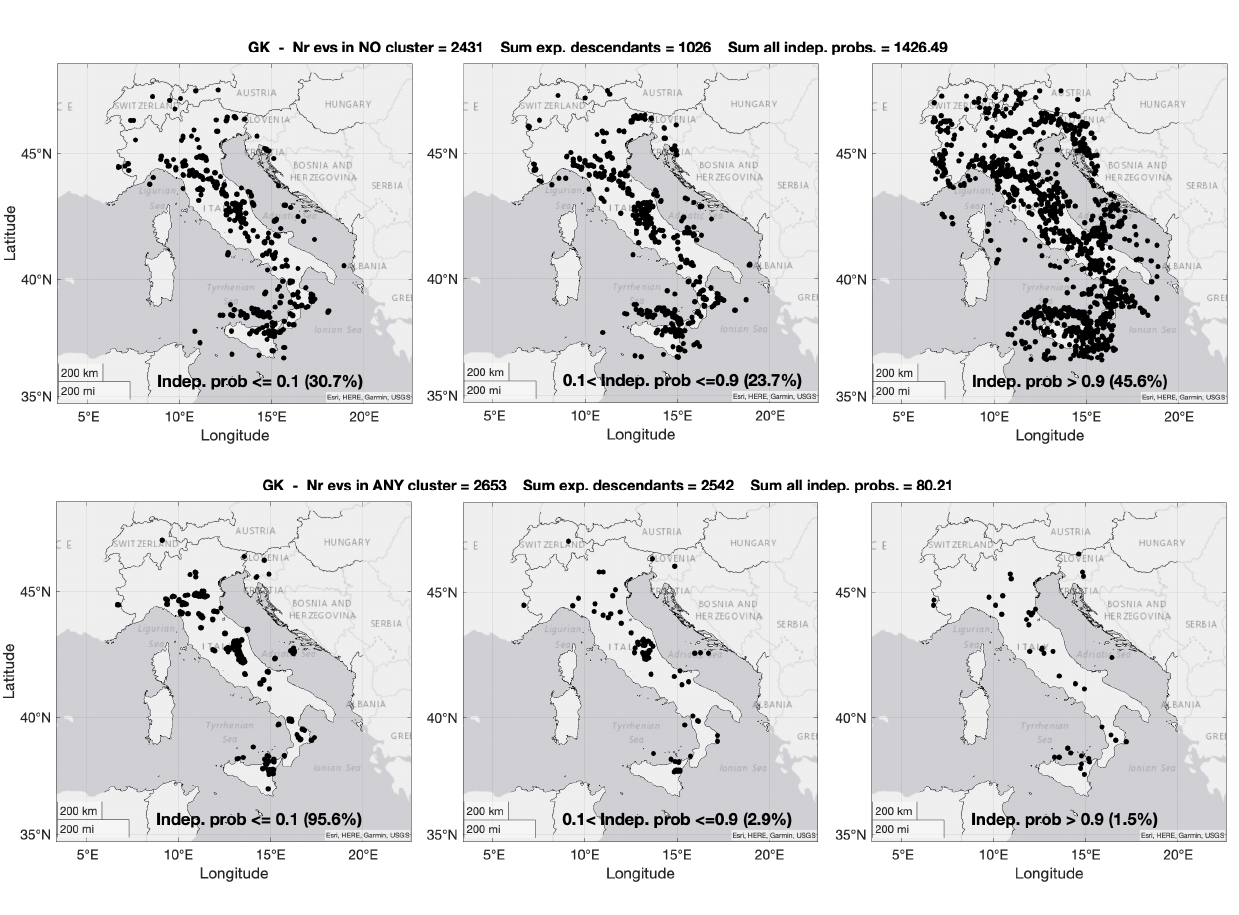}
    \caption{Seismic maps of the ISIDe events that do not belong to any GK-cluster (top) or belong to any GK-cluster (bottom). Left, middle and right panels for independence probability $\le 0.1$, in $(0.1, 0.9]$, and $>0.9$, respectively.}
    \label{fig:figure3}
\end{figure}

By looking at the independence ETAS probabilities associated with the I-not-GK events and the GK-clusterized events, respectively given in panels b) and d) of Figure~\ref{fig:figure1}, as well as in top and bottom panels of Figure~\ref{fig:figure3}, we deduce again that most of ``single'' events are either ``very likely independent'' or ``very likely triggered'', while the clustered ones are almost all ``very likely triggered''. Indeed, 76.3\% of I-not-GK events have independence probability in the intervals $[0, 0.1]$ or $(0.9, 1]$, while 95.6\% of GK-clusterized events have independence probability in $[0, 0.1]$ (Figure~\ref{fig:figure3}). All the relative percentages, the number of events considered, and the sums of their relative expected descendants and independence ETAS probabilities, are reported in Table~\ref{tab:tableS2} of the Supplement. Overall, it is possible to see that both window-based methods (ULG and GK) agree well with ETAS if we consider events recognized as in clusters, because most data have independence probability $<0.1$. Vice-versa, if we consider events not in clusters according to ULG and GK, the methods' response does not agree, because 1/3 of them still has an independence probability $<0.1$. As before, a focus on the spatio-temporal location of the I-not-GK events with independence probability in the $[0, 0.1]$ bin is shown in Figure~\ref{fig:figureS4} of the Supplement, bottom panel: events are coloured according to their occurrence year and have size increasing with the magnitude (from ML 2.9 to ML 5.4). The two panels in this figure, relative to I-not-ULG and I-not-GK events, are very similar, with main differences for offshore events, where there may be problems connected with the location accuracy.  The FMDs of the ``single'' and clusterized events in the GK-case are also given in top right and middle right panels of Figure~\ref{fig:figureS5} in the Supplement, respectively. Results are totally consistent with those obtained for the ULG-case. 

\begin{figure}
    \hspace{-2.75cm}
    \includegraphics[trim=4cm 2cm 5cm 2cm,clip,width=21cm]{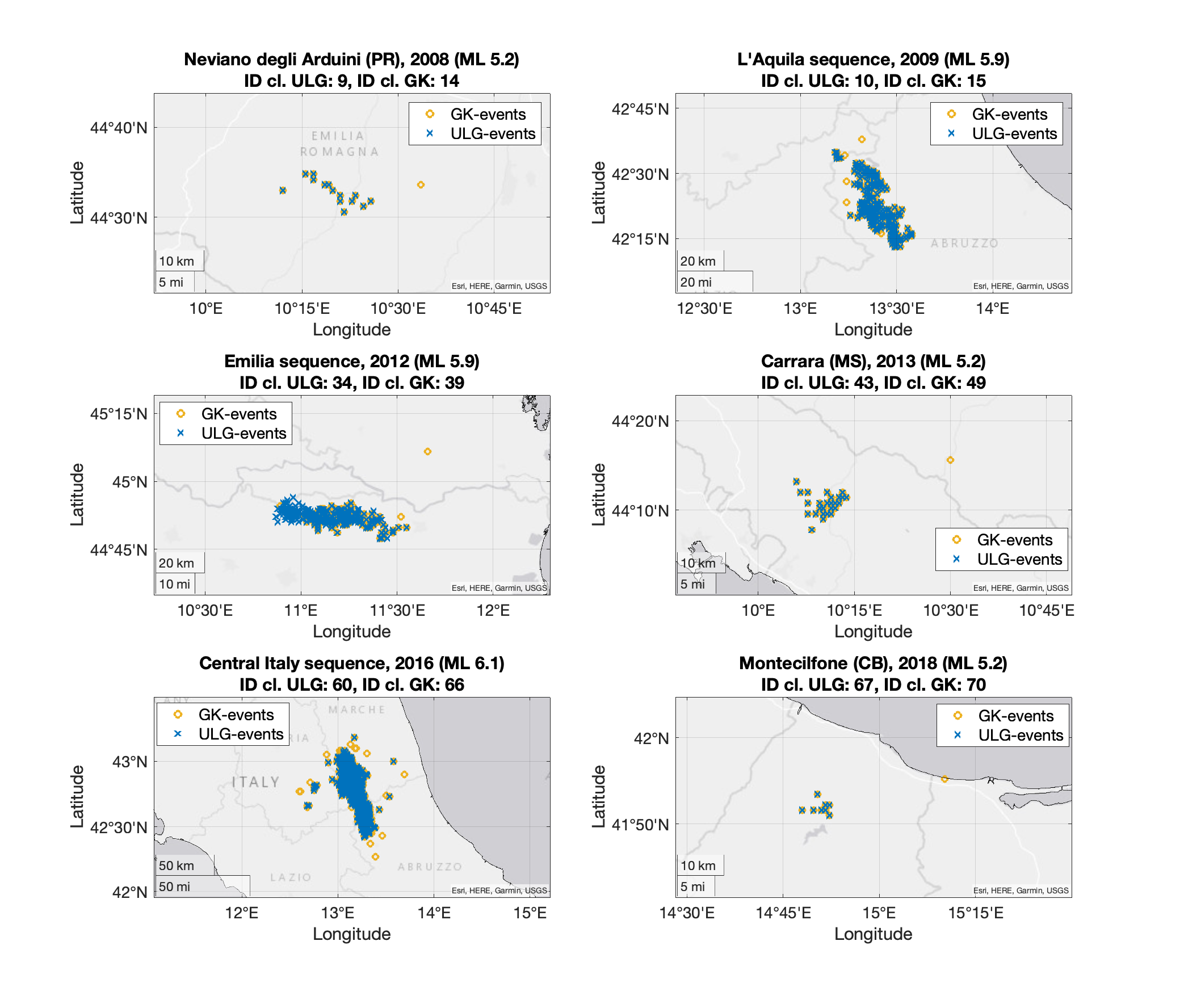}
    \caption{Seismic maps of the ULG- and GK-clusters (x markers and circles, respectively) with strongest magnitude ML $>$ 5.0. Involved sequence, clusters ID and the magnitude of the strongest event is specified for each case.}
    \label{fig:figure4}
\end{figure}

\begin{figure}
    \hspace{-1.0cm}
    \includegraphics[trim=0.01cm 0.01cm 0.01cm 0.01cm,clip,width=18cm]{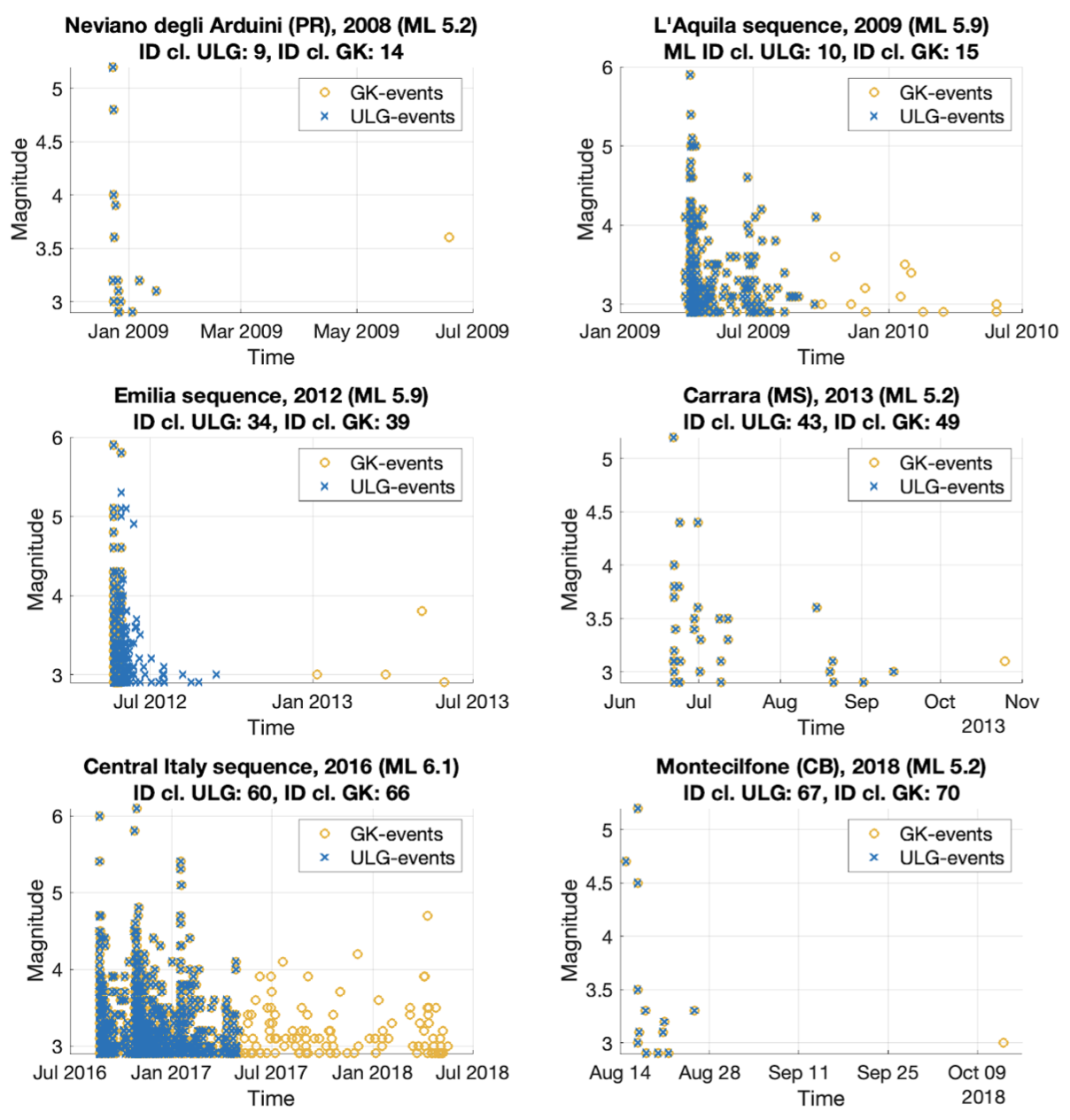}
    \caption{Magnitude VS time plot of the STR-ULG- and STR-GK-clusters (x markers and circles, respectively) with strongest magnitude ML $>$ 5.0. Involved sequence, clusters ID and the magnitude of the strongest event are specified for each case.}
    \label{fig:figure5}
\end{figure}

For the sake of graphical clarity, in what follows we will represent some results separately for the clusters with strongest magnitude ML $>$ 5.0 (``STR-ULG-'' and ``STR-GK-clusters'') and for the other clusters. We found precisely 6 STR-ULG-clusters and 7 STR-GK-clusters, whose details are reported in the Appendix~\ref{sec:appc}. Their comparison is given in Figures~\ref{fig:figure4} and~\ref{fig:figure5}. Since, as specified in this Appendix, the 2012 Emilia sequence belongs to two separate STR-GK-clusters, but to a single STR-ULG-cluster, we represent in these two figures the (longest) one which contains the ML 5.9 event. By visual inspection of Figures~\ref{fig:figure4} and~\ref{fig:figure5}, we can see that, in space, the GK procedure associates quite sparse events and produces larger clusters with respect to the ULG method. On the other hand, the ULG method creates narrower but also shorter-in-time clusters, as expected from Figure~\ref{fig:figureS2} of the Supplement, in fact, here we considered magnitudes in the interval $(4.8,6.3)$. At glance, it seems that, for larger mainshock magnitude clusters, the ULG method fails in assigning to the same cluster events close in time, thus resulting in sharply cut clusters not following the gradual decrease expected for the aftershocks of strong sequences, like those investigated here.

\section{Compare the identified clusters to the ETAS probabilities}
The previous procedures allowed us to identify a total of $N=79$ ULG-clusters and $N=82$ GK-clusters above the completeness magnitude ML 2.9, with minimum threshold for mainshock magnitude equal to 4.0. In order to assess their consistency with the ETAS approach, as explained before, we now consider the events in each of the $n^{th}\, (n=1,..,N)$ cluster and, for all of them, we trace back to both: the corresponding probability of being independent, and the expected number of triggered events according to the ETAS model, as computed through the approach by \cite{console:2010}.
 
The assessment procedure we consider is based on two simple statements. Let us assume that the $n^{th}$ cluster (nCL) contains NR events. In order for this nCL cluster to be consistent with the ETAS model, we should verify the following two checks, the first of which is also statistically tested.
\newline \textbf{- TEST 1:}
\newline the sum S1 of the expected numbers of events triggered by the NR events in the current nCL cluster should be close to the number of elements in nCL, i.e. TEST 1 $=$ S1/NR $\sim$ 1. This is because the expected offspring in the current cluster should reflect its cardinality. If we find that S1$>$NR, we can say we are in the case of an ``over-productive'' seismic sequence included in the cluster. This could be ascribed to strong sequences.
\newline \textbf{- CHECK 2:}
\newline the sum S2 of the independence probabilities of all the NR events in the current nCL cluster should be close to 1, i.e. CHECK 2 = $|$S2-1$|$ $\sim$ 0. This is because we expect a ``single'' cluster to have a ``single'' independent event.  If we find that S2 $>$ 1, we can say that the current nCL cluster has more than one independent event, and this could be the case of clusters involving very strong seismic sequences, typically characterized by several strong events which ETAS may label as independent.

We now can proceed to compare the ULG- and GK-clusters, obtained with the respective procedures, to the ETAS probabilities.

\begin{figure}
    \centering   
    \includegraphics[trim=0.01cm 0.05cm 0.01cm 0.05cm,clip,width=16cm]{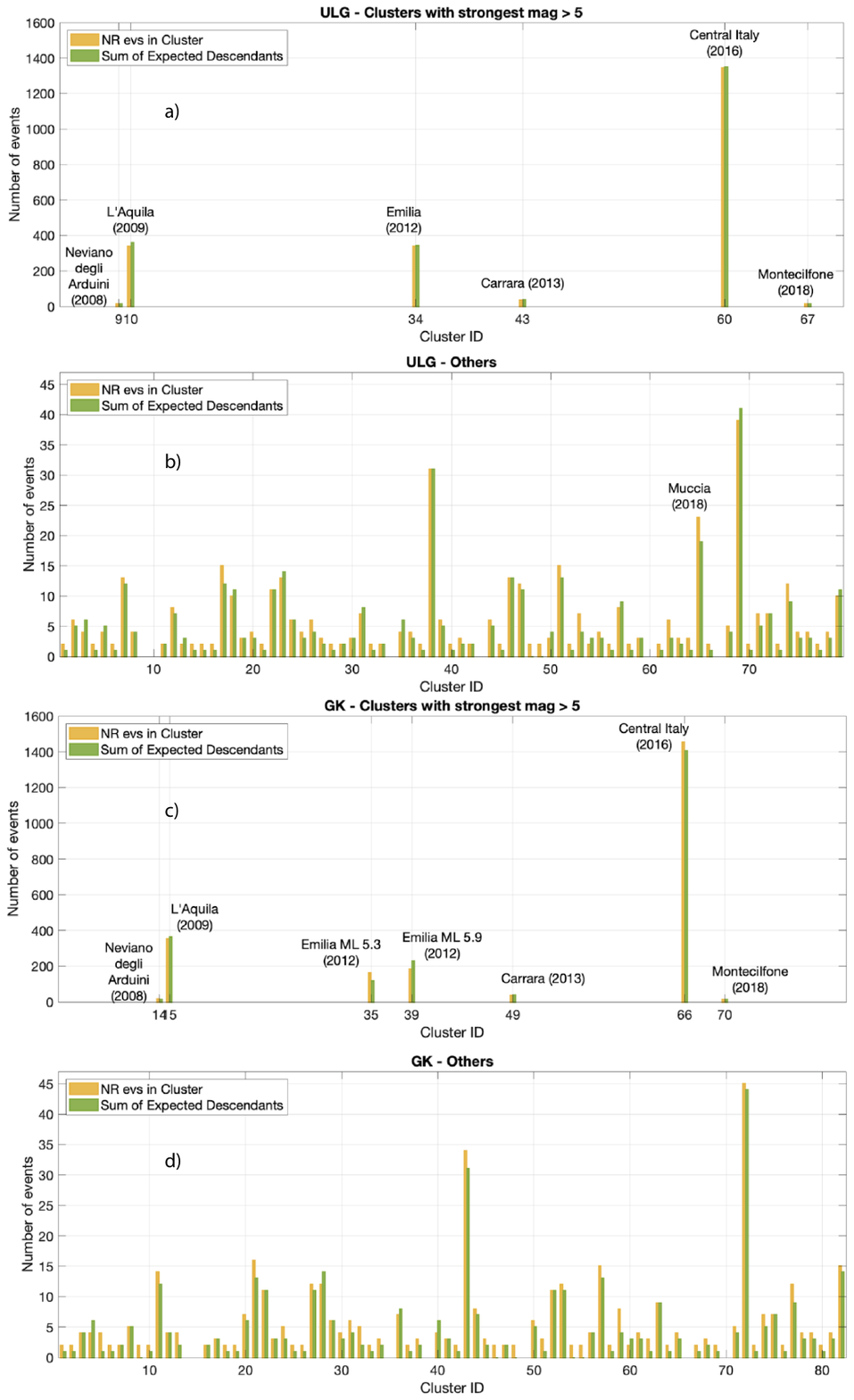}
    \caption{Histograms relative to the number of events and the sum S1 for each cluster. Panels a) and c) concerns the clusters with the strongest event having ML $>$ 5.0, respectively for the ULG- and the GK-case; panels b) and d) concerns all the other clusters, again respectively for the ULG- and the GK-case.}
    \label{fig:figure6}
\end{figure}

The numbers of events in each cluster, versus the corresponding sum S1 of expected descendants, are given in Figure~\ref{fig:figure6}, respectively panels a) and b) for ULG-clusters, and panels c) and d) for GK-clusters. As anticipated in the previous section, for the sake of simplicity we represent here the STR-clusters, and all the others, separately (panels a) and c) for ULG-STR and GK-STR; panels b) and d) for the remaining ULG- and GK-). The largest differences between the two numbers for the ULG-clusters are observed for the cluster with L'Aquila sequence (2009). L'Aquila represents one of the strongest sequences experienced in Italy in the last decades, which entailed a strong incompleteness into the catalog. Although smaller, a difference can be appreciated also for the clusters with the Central Italy sequence (2016), and the cluster with ID 65. As specified in the Appendix, the Central Italy sequence started with an ML 6.0 ($M_w$ 6.0) event that occurred in Accumoli (Rieti province, Lazio region) on 2016-08-24; this event was followed by other 3 strong events: on 2016-10-26 with ML 5.9 ($M_w$ 5.9) in Visso (Macerata province, Marche region), on 2016-10-30 with ML 6.1 ($M_w$ 6.5) in Norcia (Perugia province, Umbria region) and on 2017-01-18 with ML 5.4 ($M_w$ 5.5) in Capitignano (L'Aquila province, Abruzzo region); this is the strongest sequence recorded in the last decades in Italy. As regards the cluster with ID 65, it contains instead the moderate Muccia sequence (Macerata province, Marche region), that occurred in April 2018, with the strongest event on 2018-04-10  03:11:30 (UTC) having magnitude ML 4.7 ($M_w$ 4.6). This sequence is related to the Central Italy one, which indeed was characterized by a strong seismic activity that extended for several years. In the case of the GK procedure, the Muccia sequence (2018) is in fact included in the GK-cluster with the Central Italy sequence, for which the largest difference is observed. A similar large difference is observed also for the two GK-clusters containing the two strongest events (ML 5.9 and ML 5.3) of the Emilia sequence (2012). We recall that the GK procedure associates these events to separate clusters. 

\begin{figure}
    \centering   
    \includegraphics[trim=13cm 5cm 11cm 8cm,clip,width=13cm]{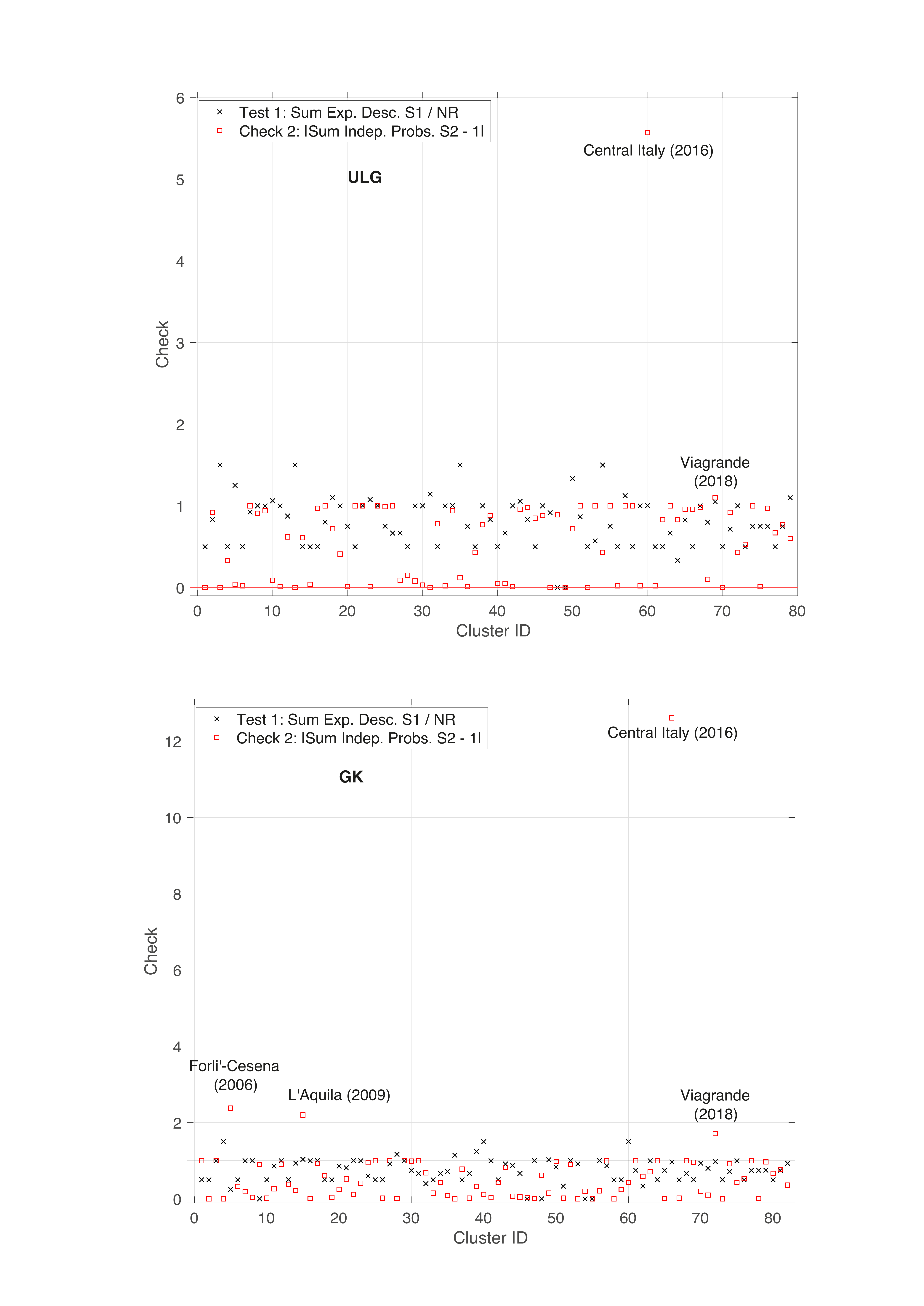}
    \caption{Graphical results of the two checks (TEST 1, x markers; CHECK 2, circles) for the ULG- and the GK-clusters in top and bottom panels, respectively.}
    \label{fig:figure7}
\end{figure}

\begin{figure}
    \hspace{-2.85cm}
    \includegraphics[trim=0.1cm 0.01cm 0.01cm 0.01cm,clip,width=21cm]{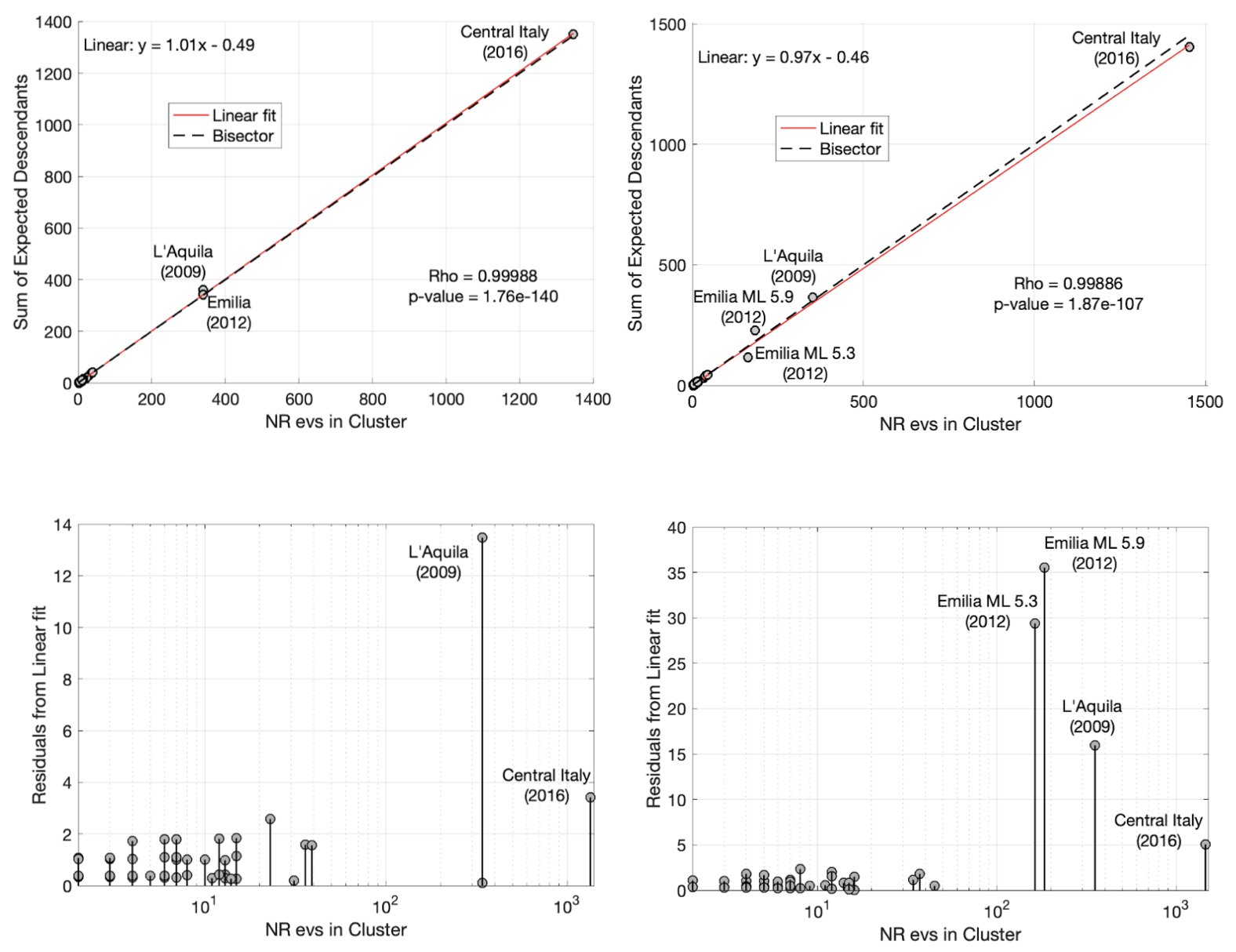}
    \caption{Correlation analysis between the sum S1 of expected descendants and the number of events in the cluster, in support of TEST 1 results, relative to the ULG- and GK-clusters, respectively in left and right panels. Top panel shows the linear fit of the data (dashed line, equation line written in top left), the comparison with the bisector (continuous line) and the results of the correlation test (bottom right). Bottom panel shows instead the residuals of the data with respect to the linear fit, in logarithmic x-scale.}
    \label{fig:figure8}
\end{figure}

In Figure~\ref{fig:figure7} we show instead the results of the two checks (TEST 1, x markers; CHECK 2, circles), top and bottom panels respectively for ULG- and GK-clusters. Comparing these two panels we can see that, while for most ULG-clusters TEST 1 ranges between 0.5 and 1.5 (Mean $=$ 0.805, Median $=$ 0.8, Standard deviation (Std) $=$ 0.308), for GK-clusters TEST 1 is generally $\le$ 1 and most GK-clusters range between 0.5 and 1 (Mean $=$ 0.737, Median $=$ 0.75, Std $=$ 0.324), showing that this algorithm tends to systematically slightly overestimate the number of events in the clusters expected by ETAS. However, we can say that the consistency of both the ULG- and GK-clusters with ETAS as regards TEST 1 is statistically significant, in fact, S1 is highly positively correlated with NR: a correlation test returned a p-value = 1.76e-140 and a p-value = 1.87e-107 for ULG- and GK-cases, respectively, both much lower than the significance level 0.05, thus implying the rejection of the ``no correlation'' null hypothesis. This is shown in the top panels of Figure~\ref{fig:figure8} (ULG- and GK-cases, respectively in left and right columns), where a linear fit is also shown (very close to the bisector). The residuals with respect to it are given in the bottom panels of the same figure. Top panels in this figure show the presence of some high leverage points, which indeed correspond to be the clusters with the Central Italy and L'Aquila sequences for the ULG-case, to which are also added the clusters with the Emilia sequence in the GK-case. For these clusters, the leverage is specifically higher than the threshold $3(k+1)/N$, where $k+1=2$ is the number of parameters in the linear model, and $N$ is the total number of clusters (79 and 82 for ULG- and GK-cases). We stress that, even if by looking at the top left panel of Figure~\ref{fig:figure8} the ULG cluster containing l'Emilia sequence seems to be far from the others, the corresponding computed leverage is lower than the threshold, and therefore this ULG-cluster cannot be ascribed as a high leverage one. In order to investigate the influence of the high leverage clusters, we excluded them from the datasets, and repeated the same correlation and linear regression analysis as before. Regarding the ULG-clusters, the linear fit models with and without high leverage clusters are very close: $1.007 x - 0.49$ and $1.009 x - 0.69$, respectively. We obtained also: (i) an R2 of about 99.9\% for both the cases with and without high leverage clusters, (ii) very close values for the standard errors relative the coefficient of the linear model (1.8e-3 and 3.5e-3 respectively for the cases with and without high leverage clusters) and (iii) both p-values, to test the null hypothesis of a 0 slope model, being much lower than the significance level of 0.05. We can therefore conclude that high leverage ULG-clusters do not influence the analysis performed: the sum of expected descendants and the number of events in clusters are indeed two correlated variables as regards ULG. In the case of GK, the linear fit models with and without high leverage clusters are $0.97 x - 0.46$ and $0.99 x - 0.91$, respectively, again with no substantial difference. The R2 value decreases from 99.7\% to 97.9\% when excluding the high leverage GK-clusters, but still indicates a very strong relation. The standard error increases from 5.2e-3 to 1.6e-2, but in both cases is small. Finally, testing the fit with a 0 slope model returns p-values $<<$ 0.05 both with and without high leverage clusters. As in the ULG-case, we can deduce that the high leverage GK clusters do not influence the correlation analysis, and there exists a significant high correlation between the sum of expected descendants and the number of events in the GK-clusters.

\begin{figure}
    \centering   
    \includegraphics[trim=0.1cm 0.05cm 0.1cm 0.05cm,clip,width=15cm]{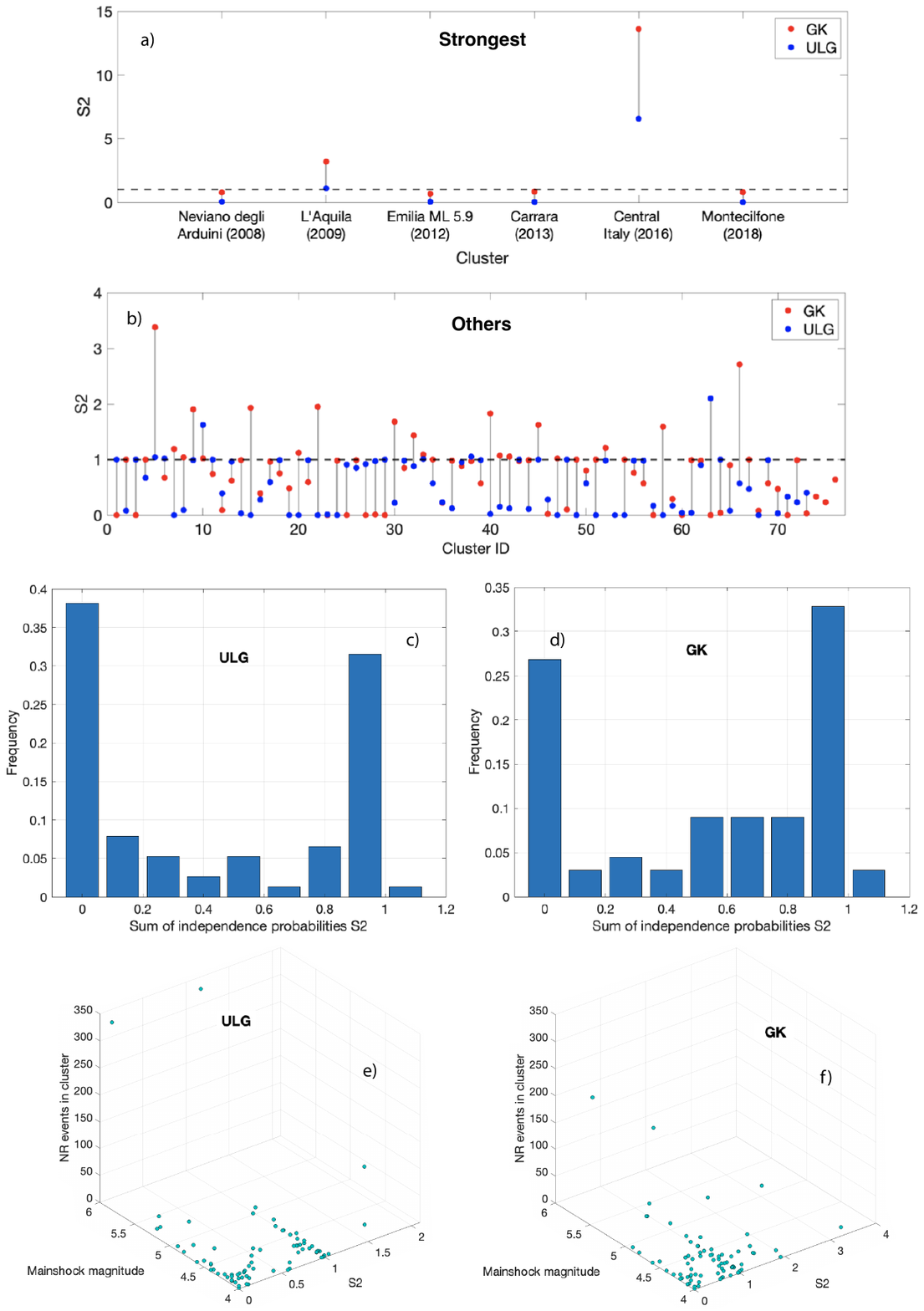}
    \caption{Panels a) and b), respectively for STR-clusters and all the others, show the comparison between the sums S2 of the independence probabilities relative to GK-clusters (red points) and ULG-clusters (blue points). Panels c) and d) show instead the histograms of the sum of independence probabilities S2, respectively for ULG- and GK-clusters, with S2 $\le$ 1.1. Panels e) and f) finally show the 3D scatter plots of the number of events in each cluster versus the independence probability S2 (x-axis) and the clusters' mainshocks' magnitude (y-axis), respectively for the ULG- and the GK-cases; we excluded here the clusters with number of events larger than 400 to avoid outliers (one ULG-cluster and one GK-cluster).}
    \label{fig:figure9}
\end{figure}

The majority of both ULG- and GK-clusters have CHECK 2 ranging between 0 and 1. A high percentage of ULG-clusters have S2 $<1$, while, in the GK case, there are 21 clusters with S2 $>1$: this is visible in panels a) and b) of Figure~\ref{fig:figure9} (respectively for STR-clusters and the others, and red/blue for GK-/ULG-cases). This latter fact is likely related to the higher number of events in the GK-clusters, which are larger and longer than for UGL, thus increasing the value of S2. Mean, Median and Std for the quantity S2 in the case of ULG-clusters are 0.594, 0.4 and 0.833, respectively. The same quantities for the GK-clusters are instead 0.978, 0.833, 1.568. The large value of Std in GK outlines a very variable behavior from one cluster to the other. In the ULG-case we find a total of 53 (out of 79) clusters having S2 $<$ 0.1 or 0.9 $<$ S2 $<$ 1.1, while for the GK-case this total is 41 (out of 82). Panels c) and d) in Figure~\ref{fig:figure9} show the histograms of S2 respectively in the ULG- and GK-cases, when excluding the clusters with S2 $>$ 1.1. We stress that the remaining clusters constitute separate bins with very low frequency (outliers, not statistically significative), as can be deduced by looking at panels a) and b) of Figure~\ref{fig:figure9}. The highest frequencies in panels c) and d) of the same figure are observed around 1, and this is what we wanted to obtain to prove consistency, but also around 0. This latter result may be due to the fact, already specified above, that we do not consider foreshocks in order to avoid possible multiple assignments of the same events to clusters close in space and time. It can therefore happen that a strong earthquake, considered as a foreshock, generates the events included in a cluster but does not belong to this cluster. Consequently, the included events have low ETAS independence probabilities, and the relative sum S2 is close to 0; this situation can also occur when there is more than one foreshock, all of which having small to moderate magnitude. The same result can also be obtained in the cases where the identification procedure separates, in two different clusters, some events that can be ascribed to the same mainshock -this latter being included in just one of the two clusters because we avoid multiple assignments-, but have a spatial or temporal distance just above the sharp cutoffs imposed by the window-based methods, likely with the presence of a sufficiently strong aftershock belonging to the other identified cluster. In panels e) and f) of Figure~\ref{fig:figure9} we finally show the 3D scatter plots of the number of events in each cluster, versus both the independence probability S2 and the clusters' mainshocks' magnitude, respectively for the ULG- and the GK-cases. To obtain this result, we excluded again the outliers, which are now the clusters with a number of events larger than 400 (only one ULG-cluster and one GK-cluster). There is no clear correlation between the three considered variables, except for a very slight tendency of clusters with a higher cardinality to have a smaller S2. Besides, the higher the mainshocks' magnitude, the more is clear that the great majority of S2 values are either very close to 0, or very close to 1. 

\begin{figure}
    \centering   
    \includegraphics[trim=0.1cm 0.05cm 0.1cm 0.05cm,clip,width=15cm]{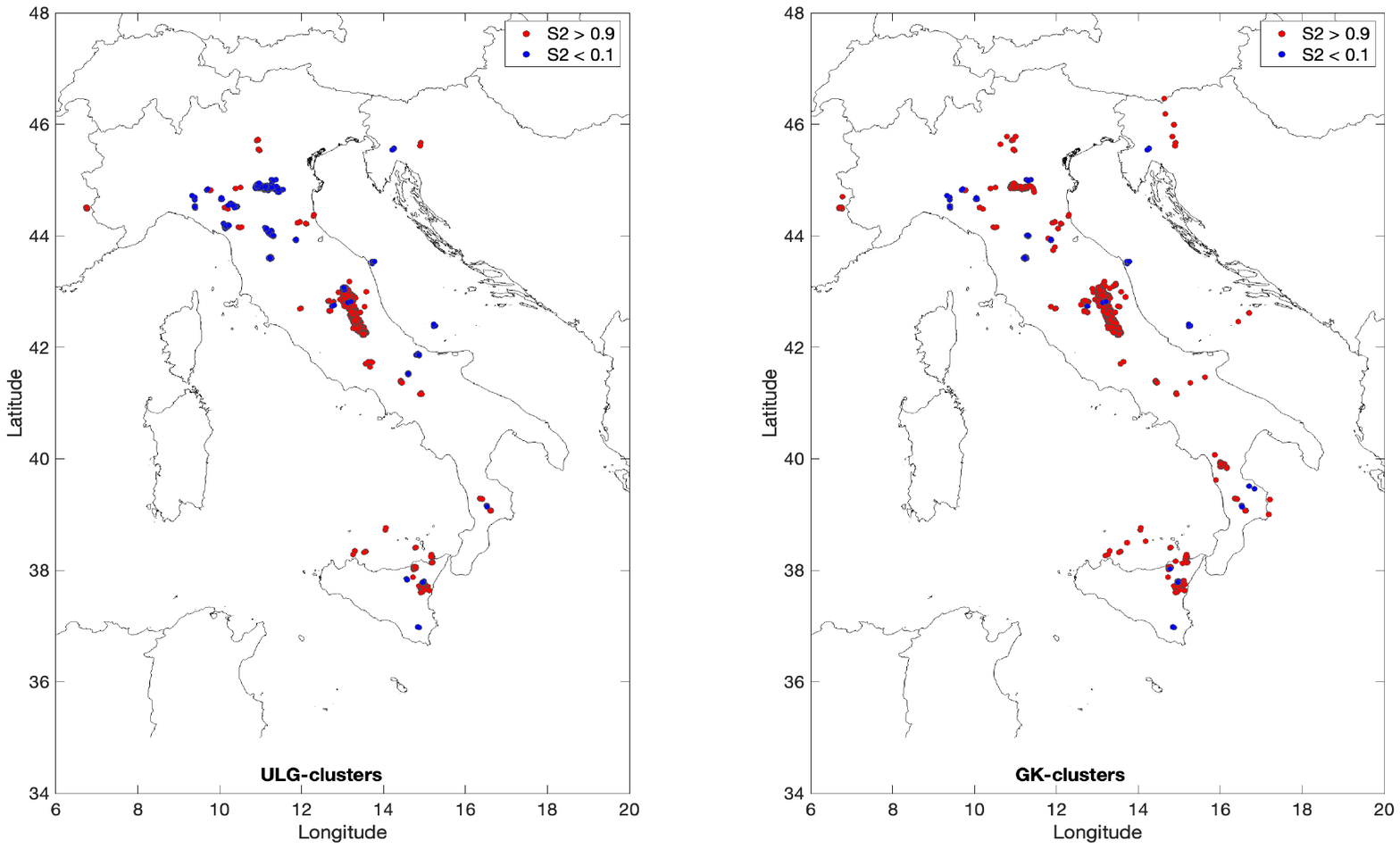}
    \caption{Seismic maps of clusters' events with S2 $>$ 0.9 (red points) and S2 $<$ 0.1 (blue points), left and right panels respectively for ULG- and GK-cases. }
    \label{fig:figure10}
\end{figure}

We eventually tried to investigate if S2 follows a given spatial pattern. Seismic maps of clusters' events with S2 $>$ 0.9 and S2 $<$ 0.1 are given in Figure~\ref{fig:figure10}, left and right panels for ULG- and GK-cases, respectively. For both these cases, a high density of S2 larger than 0.9 can be observed in the Central Apennines, while S2 $<$ 0.1 seems to stand out in the Northern part of this Mountain Chain. Interestingly, an opposite S2 behavior is observed in the Emiliana Po valley for ULG- and GK-clusters, the latter showing S2 $>$ 0.9, the former S2 $<$ 0.1. The higher ETAS independence probability of the events involved in the GK case, actually agrees with the fact that, differently from ULG, the GK procedure splits the two strongets events of the 2012 Emilia sequence in separate clusters, thus resulting in more than one mainshock. Further analyses are required to better understand any possible clear correlation between the S2 values and a physical characteristic of the seismic process, to be framed also from a statistical point of view: these will be object of future studies.

\begin{figure}
    \hspace{-2.5cm}
    \includegraphics[trim=0.1cm 0.01cm 0.01cm 0.01cm,clip,width=20cm]{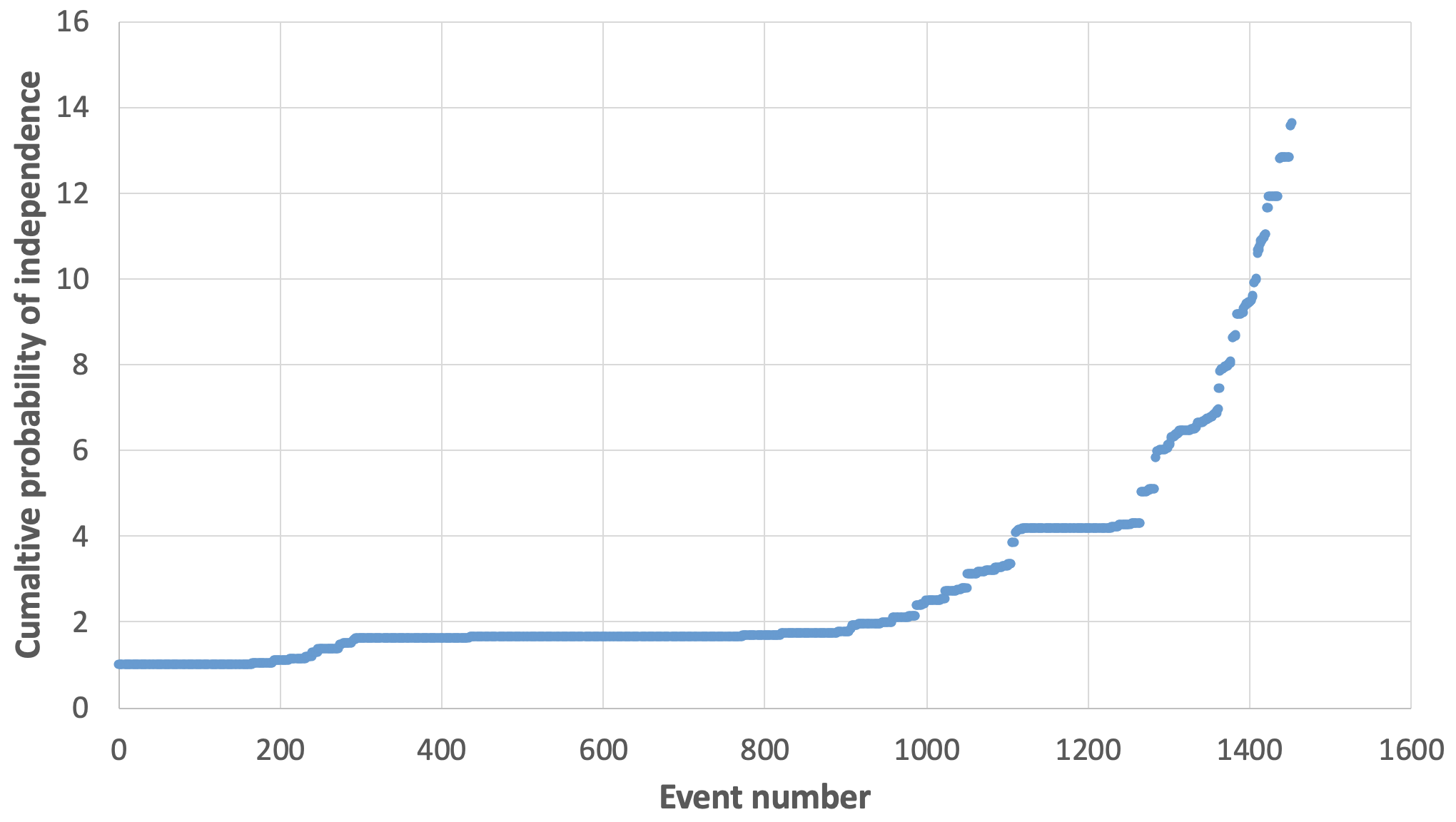}
    \caption{Cumulative ETAS probability of independence for the GK-cluster containing the Central Italy sequence (2016).}
    \label{fig:figure11}
\end{figure}

By looking again at Figure~\ref{fig:figure7}, we observe also that there are two clusters, common to the two procedures, for which the quantity $|$S2\,-1$|$ (CHECK 2) is particularly larger than 1. The first cluster is the one with the Central Italy sequence, and this is certainly due to the high productivity of the sequence itself. The second one is instead a cluster which contains the moderate Viagrande sequence (Catania province, Sicilia region) occurred in December 2018, with the strongest event on 26-12-2018 02:19:14 (UTC) with ML 4.8 ($M_w$ 4.9). The proximity of this sequence to Etna volcano allows us to ascribe it to a volcanic activity, which is known to be driven by mechanisms not well captured by the ETAS model. One may think that a high value of S2 could be also the effect of a particularly high number of ``dependent'' events, higher than those expected by the Omori law for a single sequence \citep{spassiani:2018}. However, it can be shown that in our case the real influence to S2 is given by the events with high independence probability, even if they are a few and no matter how numerous are the aftershocks. An example is given in Figure~\ref{fig:figure11}, where we plot the cumulative independence probability of the GK-cluster containing the Central Italy sequence. The first event in this cluster has an independence probability close to 1, i.e., it is a ``real mainshock''. After a series of events with low independence probability, including the strong shocks of October-November 2016, the sequence starts its second phase of epicenters' spatio-temporal scattering, characterized by events with high independence probability, reaching its maximum on May 2018. More than 50\% of S2 consists of the 10 strongest events, thus confirming that, in our case, a cluster with CHECK 2 larger than 1 contains more than a ``single immigrant'', e.g. independent event. The cumulative independence probability for the other STR-clusters, both for GK and ULG cases, are given in Figure~\ref{fig:figureS7} of the Supplement, where it can be observed that the sequences of L'Aquila 2009 and Emilia 2012 show a more complex behavior like in the Central Italy case, while in the remaining three STR-clusters there is just one very-likely independent event. In general, all the cases show that the clusters events' independence probability increases with the increase of the spatio-temporal distance between events.

In the case of the GK-clusters, we found two additional clusters that give a bad response in comparison with ETAS, again with respect to the second check. These are the one with L'Aquila sequence (as before, we expect this is due to the high productivity), and the cluster containing a small sequence occurring between the provinces of Forl\'{i}-Cesena (Emilia-Romagna region) and Arezzo (Toscana region) in the spring of 2006, with the strongest event occurred on 16 April 2006 21:15:02 (UTC) with ML 4.1 ($M_w$ 4.3). Since this sequence was either not so highly productive, nor characterized by very strong events (about 5 events with ML 3+ and one event with ML 4+ within a radius of 50 km from the strongest one, in the period April-August 2006), further investigations are needed to explain this result.

An interesting focus can be done on two other cases. The first one regards the Montecilfone (Campobasso province, Molise region) event that occurred on 16 August 2018 18:19:04 (UTC) with ML 5.2 ($M_w$ 5.1) - (ULG cluster ID 67 and GK cluster ID 70). The sequence to which this event is ascribed was characterized by some peculiar statistics (e.g. a number of aftershocks lower than expected from the classical laws adopted in statistical seismology; \cite{moretti:2018}). Interestingly, the ULG- and GK-clusters involved respond to the comparison analysis with ETAS quite well, mostly regarding the first check (TEST 1); in fact, the number of expected aftershocks reflects well the cardinality of the cluster. As regards the second check, we observe instead that the sum of independence probability is close to 0 in the ULG case, lower than the 1 expected, suggesting that this sequence was not characterized by a ``true mainshock'' according to ETAS. The second case that is worth focusing on, is the Pollino sequence (Cosenza province, Calabria region), that occurred in September-October 2012 with the strongest earthquake on 25-10-2012 23:05:24 (UTC) having magnitude ML 5.0 ($M_w$ 5.2) - (ULG cluster ID 38 and GK cluster ID 43). This sequence was characterized by a very productive seismic activity of moderate-to-low size, started in 2010 and lasted for about 2 years. Also in this case, the ULG- and GK-clusters involved respond very well to the first check (TEST 1), and show a quite low independence probability.

The explicit numerical results obtained for the two checks are listed in Tables~\ref{tab:tableS4} and~\ref{tab:tableS5} of the Supplement, respectively for the 22 ULG- and GK-clusters with strongest magnitude ML  4.5, rounded to the second decimal.

\section{Discussion and conclusions}
In this paper, we used an ETAS-based approach to compare two different cluster identification methods applied to the Italian ISIDe catalog from  2005 April 18, to 2021 April 30. Specifically, we consider the window-based clustering procedures by Gardner-Knopoff (GK) and Uhrammer-Lolli-Gasperini (ULG). We associate to all the events in the identified clusters the probability of being independent and the expected number of aftershocks derived from the ETAS model. Finally, we compare the deterministic and probabilistic approaches applied to the same dataset by checking the consistency between the clusterized events and these stochastic quantities.

No substantial differences can be observed by comparing the two cluster identification procedures. There is obviously a difference in spatial and temporal extension due to the specific different set of equations used to determine clusters' extension, but, in general, the cardinality and mainshocks of the corresponding identified clusters are comparable. This is shown for the clusters with strongest magnitude ML $\ge$ 4.5, given in Tables~\ref{tab:tableS1} and~\ref{tab:tableS3} of the Supplement. The only difference can be appreciated in the fact that GK identifies quite longer and quite wider clusters when considering strong mainshocks. This is what was expected, as shown in Figure~\ref{fig:figureS2} of the Supplement, and considering that the maximum magnitude in our catalog is ML 6.1. 

As regards the comparison between the deterministic window-based approach and the probabilistic ETAS, although the two clustering procedures are rather subjective, our analysis proved an overall consistency between all the identified clusters and the relative events' ETAS probabilities. There are only small incoherencies, which are however legitimated by the fact that the two approaches are based on two completely different grounds (deterministic vs probabilistic). Necessarily, the window-based method imposes a sharp cutoff to include events or not in a cluster, differently from the probabilistic quantities associated to the events by ETAS. In particular, as regards the CHECK 2 (that is, the sum S2 of the independence probabilities within a cluster being or not close to 1), we have shown that the independence probability tends to increase in the final part of a sequence (higher spatio-temporal distances between events) and that, regardless of how many they are, the aftershocks do not give a relevant contribution to S2. The identified clusters' response to CHECK 2, and specifically the fact that ETAS recognizes more than a single independent event in some clusters (S2 $>$ 1), may reflect that the window-based identification procedures requires the labeling of an event as a mainshock to be performed, while ETAS does not account for such labeling. A high density of cases with S2 $>$ 0.9 can be finally observed in the Central Apennines.

It is also important to stress that it is not always easy to distinguish between sequences with a single ``mainshock'' and sequences containing a high number of moderate events (swarms). In general, such distinction is better captured by a probabilistic approach such as ETAS. Similarly, a strong event following a former strong one can be associated to this latter's same cluster, or to a different cluster, depending on the cluster identification method adopted and the specific magnitudes considered (an example here is the Emilia sequence). Instead, aside from that, the ETAS model would associate to both these events a specific probability of being independent.  

In general, the deterministic and the probabilistic approaches surely allow us to pursue the analysis from two different perspectives, and highlight different aspects of seismicity. Except for selecting the stochastic model to consider, the probabilistic view is less subjective in the sense it does not take into account a specific threshold for characterizing the events, for example to be in a cluster or not. Still, probability is a challenging concept to understand and interpret, and of course, by definition, it carries a certain degree of uncertainty for every interpretation it entails. The message we would like to be taken home is that there is no general rule for one approach being preferable to the other, but it is important to be aware of the meaning behind the selected approach and the implications, in order to properly interpret the results obtained. Besides, when the probabilistic/deterministic approach adopted is correct, no substantial differences should be expected.

\section{Figures}

\clearpage

\section*{Acknowledgments}
Funded by Seismic Hazard Center - WP3 Short-Term Probabilistic Seismic Hazard (CPS, INGV). Developed within the Bilateral project Italy-Japan, funded by the Italian Ministry of Foreign Affairs and International Cooperation, and the NEar real-tiME results of Physical and StatIstical Seismology for earthquakes observations, modeling and forecasting (NEMESIS) Project (INGV).

\cleardoublepage
\addcontentsline{toc}{section}{\refname}



\clearpage

\appendix

\section{The spatio-temporal ETAS model}
\label{sec:appa}
The specific formulation of the ETAS rate density of earthquakes we consider in this paper is: 

\begin{linenomath*}
\begin{eqnarray*}
\lambda(t,x,y,m) &=& f_r \,\lambda_0(x,y,m) + \sum_{i=1}^N H(t-t_i) \lambda_i(t,x,y,m)
\end{eqnarray*}
\end{linenomath*}

where $(t,x,y,m)$ indicate temporal occurrence, spatial location and magnitude of the event, and

\begin{linenomath*}
\begin{eqnarray*}
\lambda_0(x,y,m) &=& \lambda_0(x,y) \beta e^{-\beta(m-m_0)}
\end{eqnarray*}
\end{linenomath*}

is the rate density of the background events,  $f_r$ is the fraction of spontaneous over the total number of events (failure rate), $m_0$ is the completeness magnitude, $\beta = b\ln 10$ is related to the b-value (parameter that controls the proportion of larger shocks with respect to the smaller ones), $H(\cdot)$ is the step function, and

\begin{linenomath*}
\begin{eqnarray*}
\lambda_i(t,x,y,m) &=& K\left[\frac{d_i^2}{(x-x_i)^2 + (y-y_i)^2 + d_i^2}\right]^{q}\,(t-t_i+c)^{-p}\,\beta e^{-\beta(m-m_0)}, 
\end{eqnarray*}
\end{linenomath*}

is the kernel function for the triggered seismicity, where $d_i = d_0 10^{\alpha(m_i-m_0)/2}$ and $(K, d_0, q, c, p, \alpha)$ are free, positive parameters typically estimated through the Maximum Likelihood Estimation (MLE) technique.

\section{Spatio-temporal equations defining the two window-based clustering approaches.}
\label{sec:appb}
The two specific sets of equations we consider here to identify the clusters are the following.

\begin{enumerate}
    \item We use the law by \cite{uhrhammer:1986} for space and the law by \cite{lolli:2003} for time, as successfully applied in Italy by \cite{gentili:2017}:

    \begin{linenomath*}
    \begin{align}
    \label{eqn:ulg}
    d =& e^{0.804 \cdot M_m - 1.024}, \nonumber \\
    t =& 60+60(M_m-4).
    \end{align}
    \end{linenomath*}
    
    We named this set of clusters ``ULG-clusters''.

    \item We use the widely known equations by \cite{gardner:1974}, according to which the functional form of the spatio-temporal windows is obtained as:

    \begin{linenomath*}
    \begin{align}
    \label{eqn:gk}
    d =& 10^{0.1238 \cdot M_m + 0.983}, \nonumber \\
    t =& 10^{A \cdot M_m + B},
    \end{align}
    \end{linenomath*}

    where

    \begin{linenomath*}
    \begin{eqnarray*}
    & & A = 0.032, \;\;\;B = 2.7389  \;\;\;if \;\;\; M_m \ge 6.5; \\
    & & A = 0.5409, \;B = -0.547 \;\;\; if \;\;\; M_m < 6.5.
    \end{eqnarray*}
    \end{linenomath*}

    We named this set of clusters ``GK-clusters''.

    \end{enumerate}

In both cases, $M_m$ is the mainshock magnitude, $t$ is expressed in days, $d$ in kilometers.

\section{Largest ULG- and GK-clusters and their comparison}
\label{sec:appc}
As explained in Section ``Identification of ULG- and GK-clusters, and their comparison with ETAS independence probability'', we identified 79 ULG-clusters and 82 GK-clusters. Here we will give some details relative to those containing the strongest magnitude ML $\ge$ 5.0 as, for the sake of simplicity and graphical clarity, these clusters are represented separately in several analyses discussed in Section ``Compare the identified clusters to the ETAS probabilities''. The ULG- and GK-clusters which contain the strongest magnitude ML $>$ 5.0 (``STR-ULG-'' and ``STR-GK-clusters'', respectively) are the following (for the reference of the ID clusters below, see Tables~\ref{tab:tableS1} and~\ref{tab:tableS3} in the Supplement):

\begin{enumerate}
    \item ML 5.2 ($M_w$ 4.9) in Neviano degli Arduini (Parma province, Emilia-Romagna region), 2008-12-23 sequence, corresponding to the ULG cluster ID 9 and GK cluster ID 14, this latter being about 5 months longer than the former one;

    \item L'Aquila sequence (Abruzzo region), 2009; strongest event on 2009-04-06 with ML 5.9 ($M_w$ 6.1); this sequence is in ULG cluster ID 10 and GK cluster ID 15, this latter being about 8 months longer than the former (ULG-)one; 
    
    \item Emilia sequence (Northern Italy, Emilia-Romagna region), 2012; strongest events on 2012-05-20 with ML 5.9 ($M_w$ 5.8), and on 2012-05-29 with ML 5.8 ($M_w$ 5.6), this latter followed a few hours later by another strong event with ML 5.3 ($M_w$ 5.3); in the case of ULG, this sequence is contained in a ``single'' cluster with ID 34, while the GK procedure split this sequence in two clusters with ID 35 (containing the ML 5.3 event), and ID 39 (containing the ML 5.9 and ML 5.8 events); 
    
    \item ML 5.2 ($M_w$ 5.1) in Carrara (Massa-Carrara province, Toscana region), 2013-06-21; this sequence is in ULG cluster ID 43 and GK cluster ID 49, this latter being about 1 month longer than the former one;

    \item Central Italy sequence (Lazio, Abruzzo, Toscana, Umbria and Marche regions involved), 2016; strongest events on 2016-08-24 with ML 6.0 ($M_w$ 6.0), and on 2016-10-30 with ML 6.1 ($M_w$ 6.5); this sequence is in ULG cluster ID 60 and GK cluster ID 66, this latter being about 1 year longer than the former (ULG-)one;
    
    \item ML 5.2 ($M_w$ 5.1) in Montecilfone (Campobasso province, Molise region), 2018-08-16; this event is in ULG cluster ID 67 and GK cluster ID 70, this latter being about 1.5 months longer than the former one.
\end{enumerate}

We then identified a total of 7 STR-GK-clusters and 6 STR-ULG-clusters (mapped in Figure~\ref{fig:figureS8} of the Supplement), the former being usually longer than the latter (especially in the cases of the L'Aquila and the Central Italy sequences). This is what was expected, as discussed in Section ``The two approaches used to identify clusters'' and shown in Figure~\ref{fig:figureS2} of the Supplement. Interestingly, both the ULG and GK procedures identified a cluster, containing only two events, which covers a temporal interval including the 6 April 2009, that is the occurrence day of the strongest event (ML 5.9) in L'Aquila sequence, but that does not contain this event. Indeed, both these clusters contain 2 events that occurred near Faenza (Ravenna province, Emilia Romagna region); the spatial distance justifies the reason why these events belong to a cluster different from the one including the L'Aquila sequence.

\section{Supplement}
\begin{figure}
    \centering       
    \includegraphics[trim=1cm 0.01cm 1cm 1cm,clip,width=\textwidth]{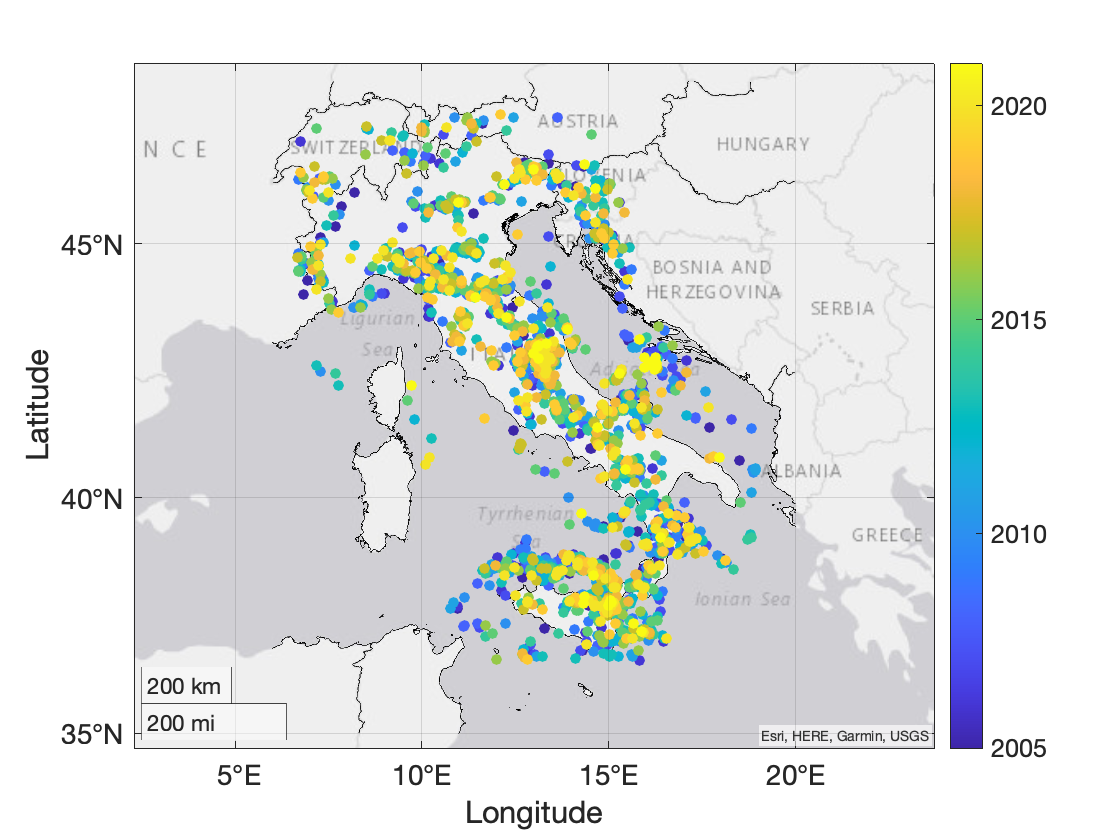}
    \caption{Seismic map of the ISIDe catalog considered for the clustering procedures. The events are coloured according to their time of occurrence.}
    \label{fig:figureS1}
\end{figure}

\noindent\textbf{Figure~\ref{fig:figureS1}.}
Seismic map of the ISIDe catalog considered for the clustering procedures. The events are coloured according to their time of occurrence.

\begin{figure}
    \hspace{-2.9cm}
    \includegraphics[trim=0.01cm 0.01cm 0.01cm 0.01cm,clip,width=20.5cm]{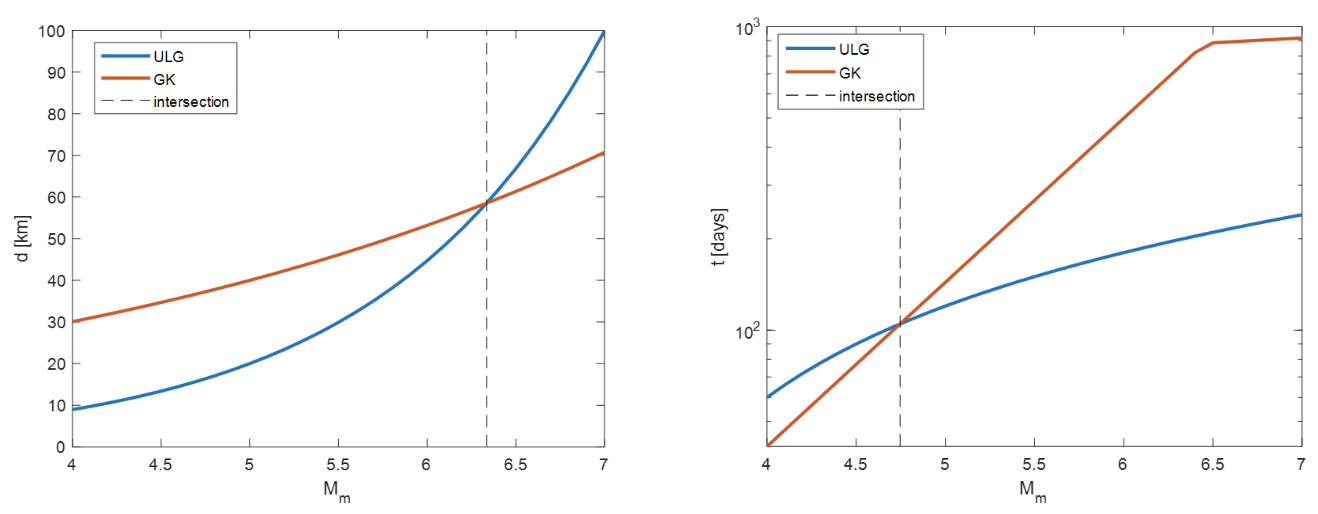}
    \caption{Comparison between the spatial (left panel) and temporal (right panel) equations by ULG- and GK-methods, for mainshocks in the considered magnitude range.}
    \label{fig:figureS2}
\end{figure}

\noindent\textbf{Figure~\ref{fig:figureS2}.} 
Comparison between the spatial (left panel) and temporal (right panel) equations by ULG- and GK-methods, for mainshocks in the considered magnitude range.

\begin{figure}
    \centering       
    \includegraphics[trim=0.01cm 0.01cm 0.01cm 0.01cm,clip,width=\textwidth]{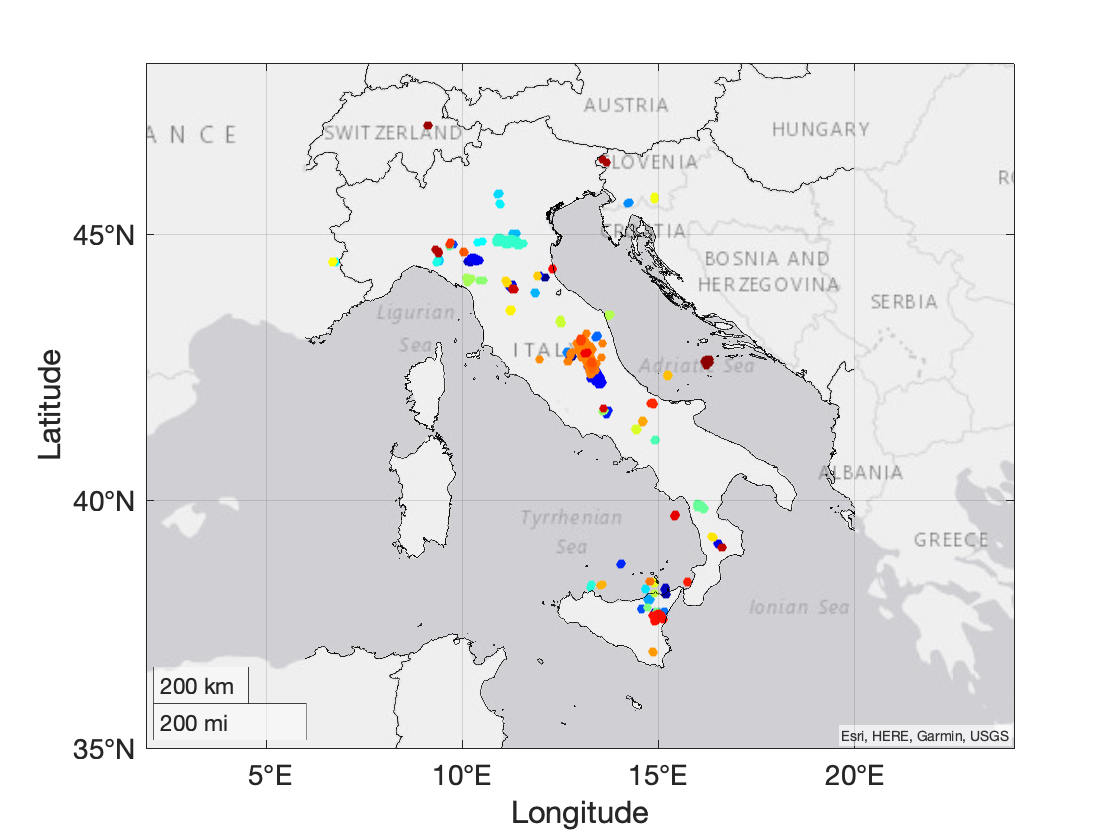}
    \caption{Seismic map of the ULG-clusters. Each cluster has a different color.}
    \label{fig:figureS3}
\end{figure}

\begin{figure}
    \centering       
    \includegraphics[trim=0.01cm 0.01cm 0.01cm 0.01cm,clip,width=\textwidth]{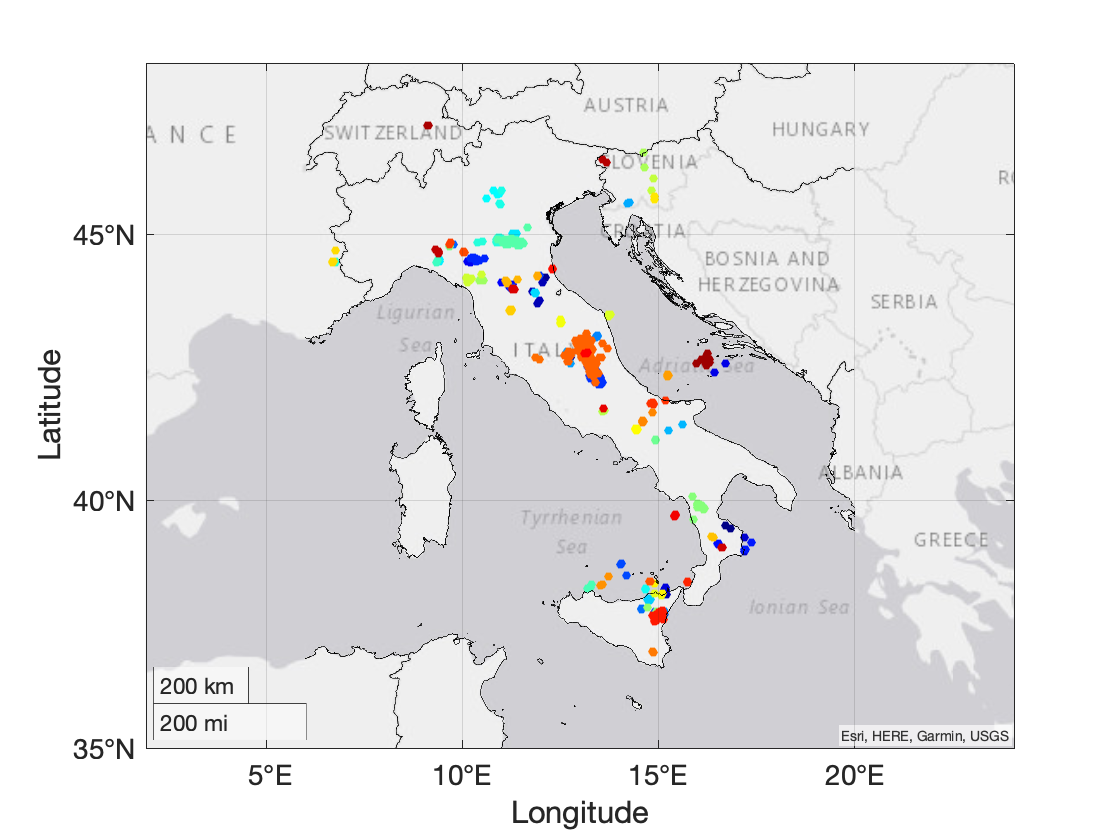}
    \caption{Seismic map of the GK-clusters. Each cluster has a different color.}
    \label{fig:figureS6}
\end{figure}

\noindent\textbf{Figures~\ref{fig:figureS3} and~\ref{fig:figureS6}.} 
Seismic map of the ULG-clusters and the GK-clusters, respectively. Each cluster has a different color.

\begin{figure}
    \centering       
    \includegraphics[trim=0.01cm 2.5cm 0.01cm 1cm,clip,width=16cm]{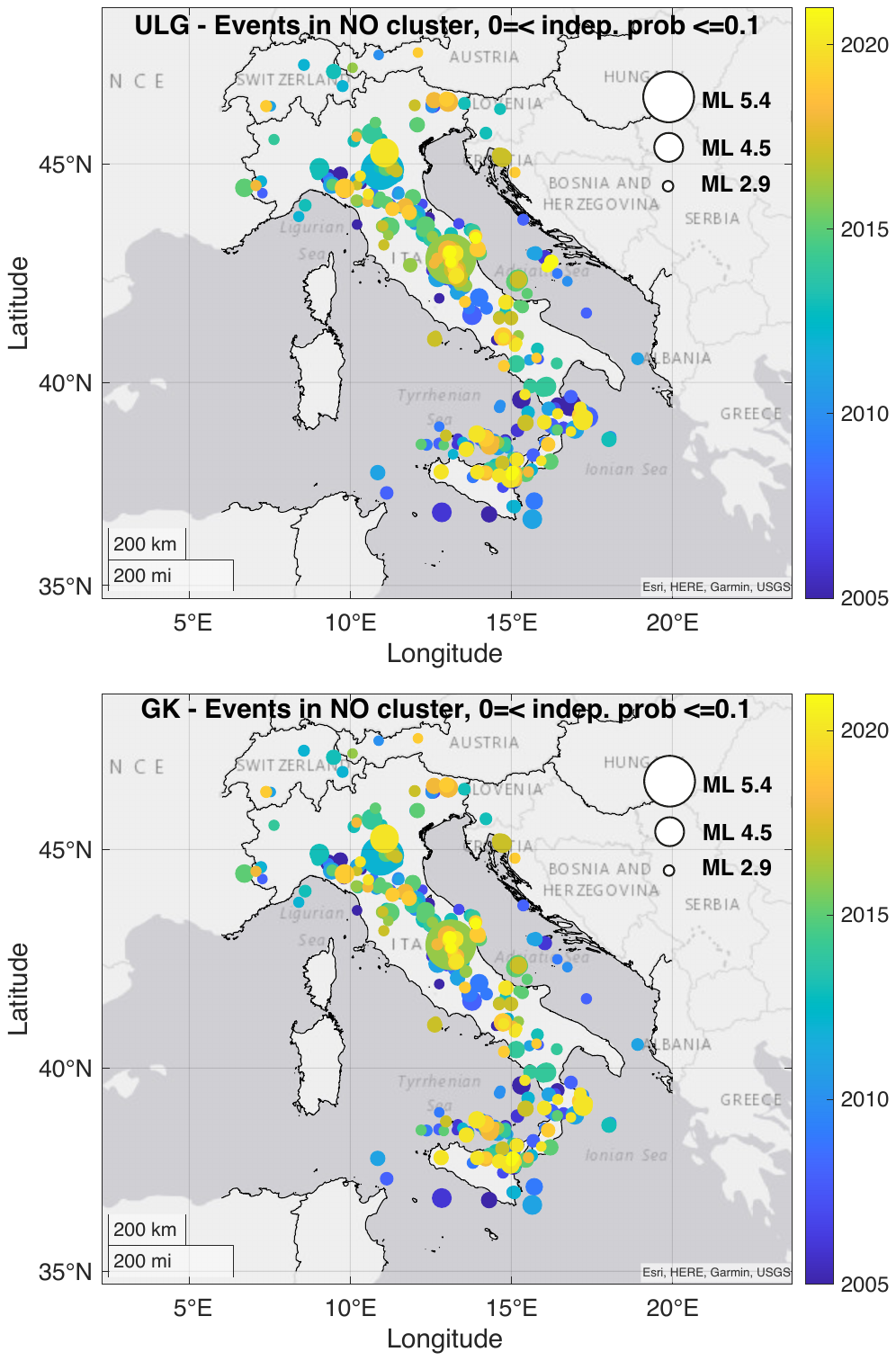}
    \caption{Seismic maps of the I-not-ULG (top) and I-not-GK (bottom) events with independence probability in the $[0, 0.1]$ bin. The events are indicated with dots that are coloured according to their occurrence year and have size that increases with the events' magnitude.}
    \label{fig:figureS4}
\end{figure}

\noindent\textbf{Figure~\ref{fig:figureS4}.} 
Seismic maps of the I-not-ULG (top) and I-not-GK (bottom) events with independence probability in the $[0, 0.1]$ bin. The events are indicated with dots that are coloured according to their occurrence year and have size that increases with the events' magnitude.

\begin{figure}
    \hspace{-0.9cm}       
    \includegraphics[trim=0.01cm 0.01cm 0.01cm 0.01cm,clip,width=16cm]{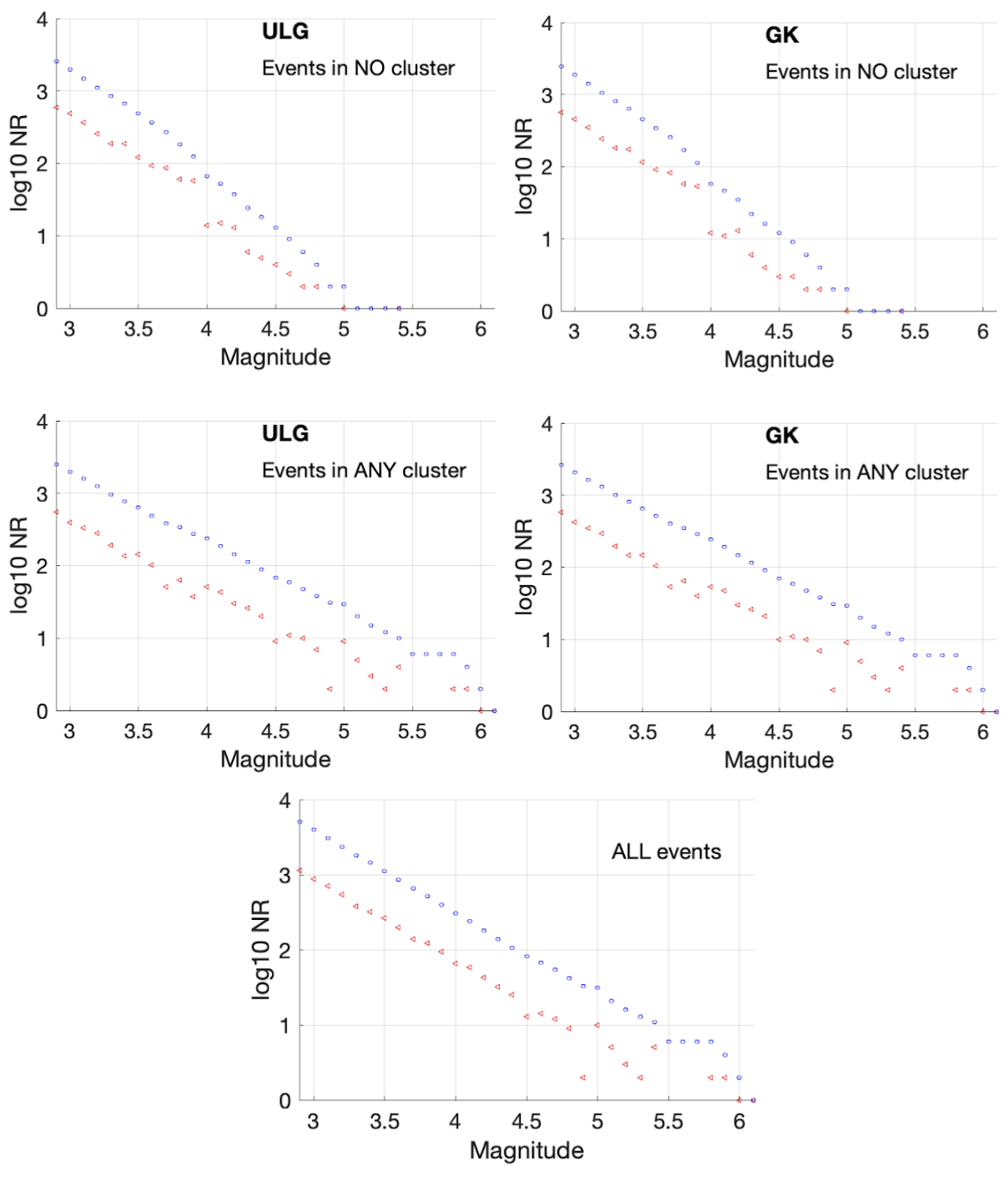}
    \caption{Probability (red triangles) and cumulative (blue circles) density distributions of the events that do not belong to any cluster, that belong to some cluster, and all the events together, respectively in top, middle and bottom line. The left (right) panels concern the ULG- (GK-) clusters.}
    \label{fig:figureS5}
\end{figure}

\noindent\textbf{Figure~\ref{fig:figureS5}.} 
Probability (red triangles) and cumulative (blue circles) density distributions of the events that do not belong to any cluster, that belong to some cluster, and all the events together, respectively in top, middle and bottom line. The left (right) panels concern the ULG- (GK-) clusters.

\begin{sidewaysfigure}
    \centering
    \hspace{-3cm}
    \includegraphics[trim=0.01cm 0.01cm 6cm 0.01cm,clip,width=\columnwidth]{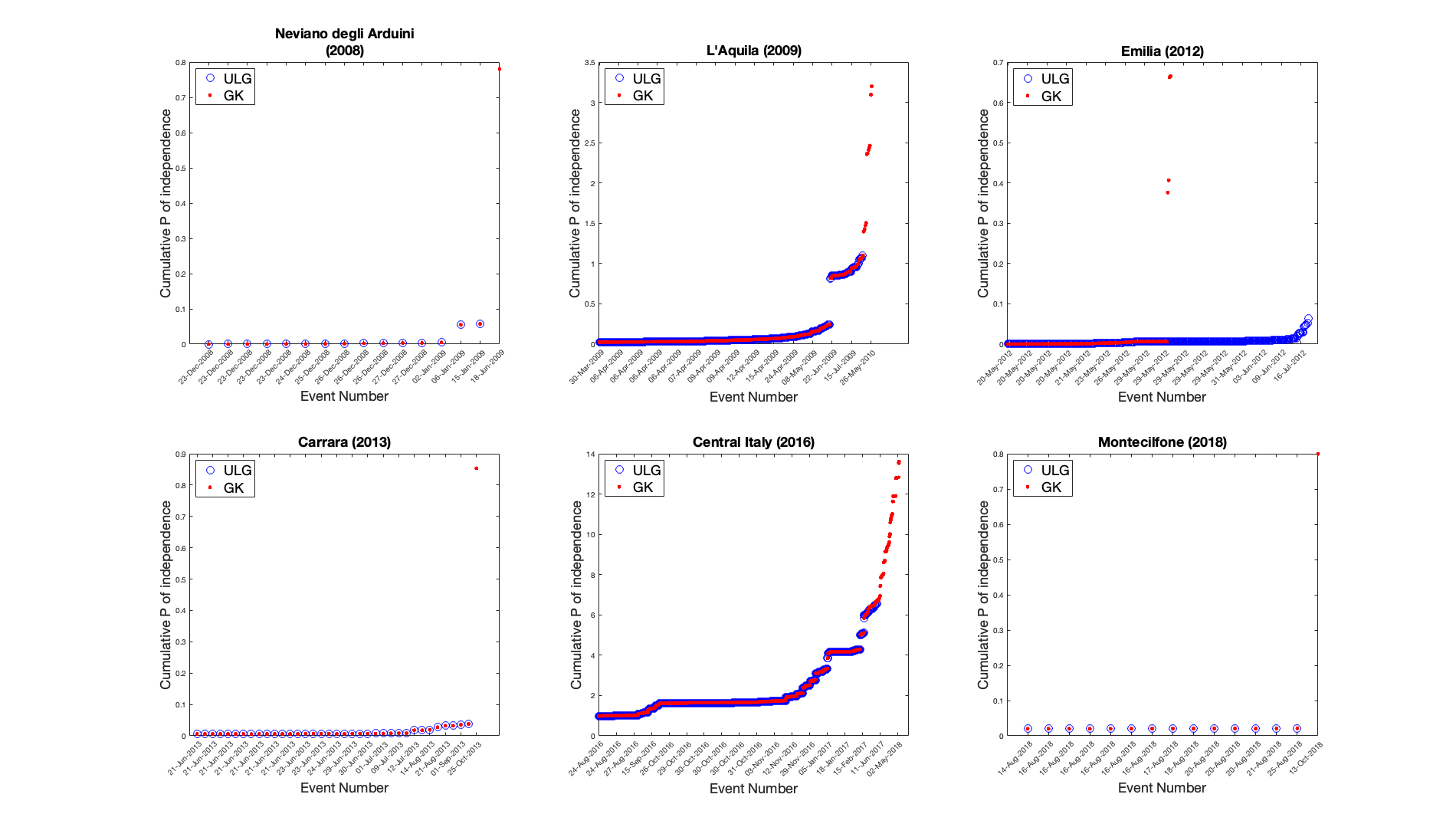}
    \caption{Cumulative ETAS probability of independence for the STR-ULG-clusters (blue circles) and the STR-GK-clusters (red points) with strongest magnitude ML $>$ 5.0.}
    \label{fig:figureS7}
\end{sidewaysfigure}

\noindent\textbf{Figure~\ref{fig:figureS7}.} 
Cumulative ETAS probability of independence for the STR-ULG-clusters (blue circles) and the STR-GK-clusters (red points) with strongest magnitude ML $>$ 5.0.

\begin{figure}
    \hspace{-3cm}
    \includegraphics[trim=0.01cm 0.01cm 6cm 0.01cm,clip,width=19cm]{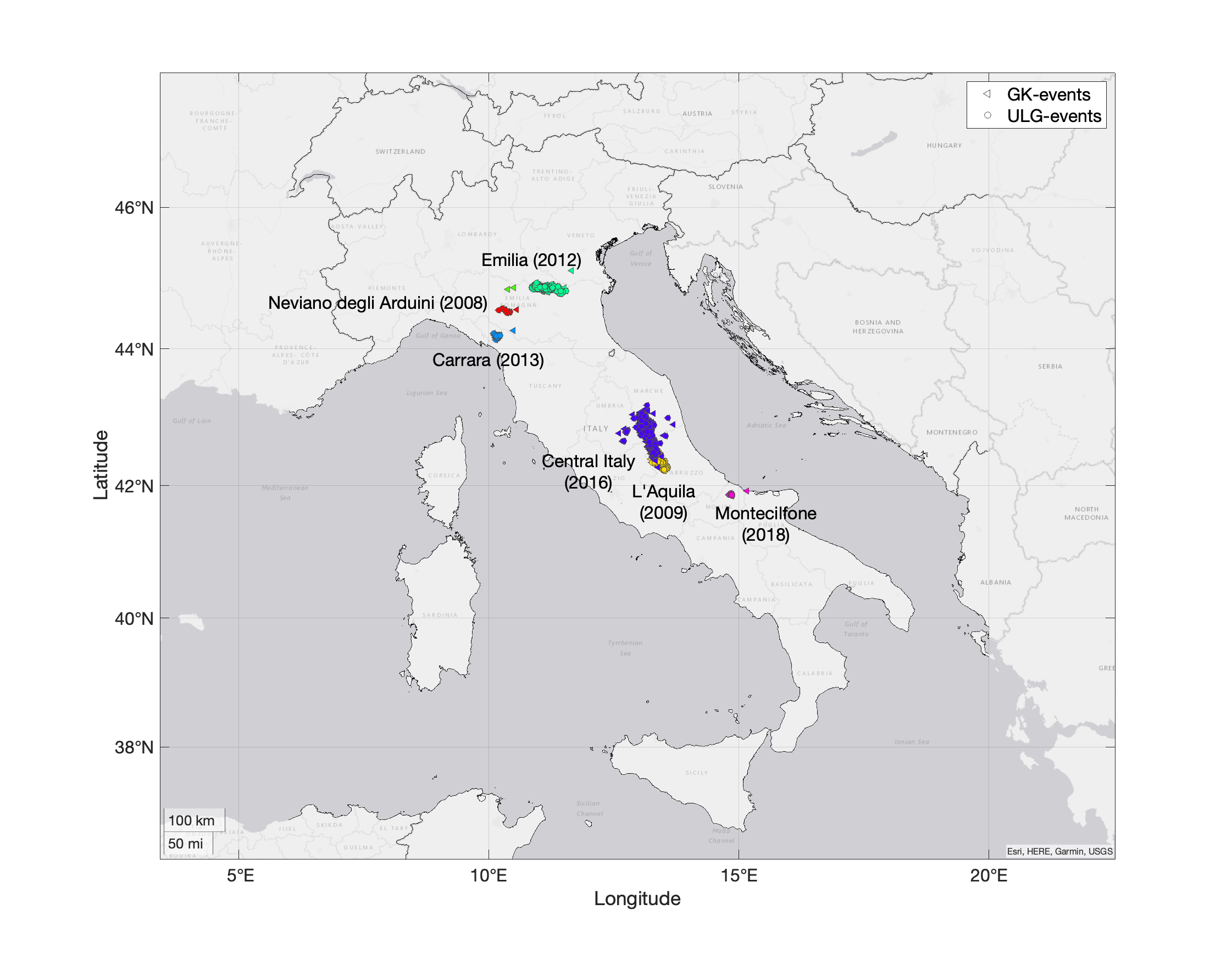}
    \caption{Seismic maps of the 7 STR-GK-clusters (triangles) and the 6 STR-ULG-clusters (circles). Each cluster has a different color. For simplicity, the 2 STR-GK-clusters containing l'Emilia sequence (2012) have the same color as that of the STR-ULG-cluster that include the same sequence.}
    \label{fig:figureS8}
\end{figure}

\noindent\textbf{Figure~\ref{fig:figureS8}.} 
Seismic maps of the 7 STR-GK-clusters (triangles) and the 6 STR-ULG-clusters (circles). Each cluster has a different color. For simplicity, the 2 STR-GK-clusters containing l'Emilia sequence (2012) have the same color as that of the STR-ULG-cluster that include the same sequence.

\noindent\textbf{Tables~\ref{tab:tableS1} and~\ref{tab:tableS3}.}
ULG-clusters and the GK-clusters, respectively, with the strongest magnitude ML $\ge$ 4.5 (UTC time).

\begin{table}[]
\caption{ULG-clusters with the strongest magnitude ML $\ge$ 4.5 (UTC time).\vspace{0.5cm}}
\label{tab:tableS1}
\hspace{-1cm}
\begin{tabular}{cccccl}
\textbf{Cluster ID} & \textbf{NR events} & \textbf{First event}                                                 & \textbf{Last event}                                                 & \textbf{Strongest event}                                            &  \\ \hline
9                   & 15                 & \begin{tabular}[c]{@{}c@{}}23-Dec-2008 15:24 \\ ML 5.2\end{tabular}  & \begin{tabular}[c]{@{}c@{}}15-Jan-2009 11:08 \\ ML 3.1\end{tabular} & \begin{tabular}[c]{@{}c@{}}23-Dec-2008 15:24 \\ ML 5.2\end{tabular} &  \\ \hline
10                  & 340                & \begin{tabular}[c]{@{}c@{}}30-Mar-2009 13:38 \\ ML 4.1\end{tabular}  & \begin{tabular}[c]{@{}c@{}}24-Sep-2009 16:14 \\ ML 4.1\end{tabular} & \begin{tabular}[c]{@{}c@{}}06-Apr-2009 01:32 \\ ML 5.9\end{tabular} &  \\ \hline
11                  & 2                  & \begin{tabular}[c]{@{}c@{}}05-Apr-2009 20:20 \\ ML 4.6\end{tabular}  & \begin{tabular}[c]{@{}c@{}}06-Apr-2009 03:33 \\ ML 3.3\end{tabular} & \begin{tabular}[c]{@{}c@{}}05-Apr-2009 20:20 \\ ML 4.6\end{tabular} &  \\ \hline
13                  & 2                  & \begin{tabular}[c]{@{}c@{}}07-Sep-2009 21:26 \\ ML 4.5\end{tabular}  & \begin{tabular}[c]{@{}c@{}}08-Sep-2009 17:19\\ ML 3.6\end{tabular}  & \begin{tabular}[c]{@{}c@{}}07-Sep-2009 21:26 \\ ML 4.5\end{tabular} &  \\ \hline
25                  & 4                  & \begin{tabular}[c]{@{}c@{}}17-Jul-2011 18:30 \\ ML 4.8\end{tabular}  & \begin{tabular}[c]{@{}c@{}}27-Jul-2011 01:13 \\ ML 3.0\end{tabular} & \begin{tabular}[c]{@{}c@{}}17-Jul-2011 18:30 \\ ML 4.8\end{tabular} &  \\ \hline
30                  & 3                  & \begin{tabular}[c]{@{}c@{}}25-Jan-2012 08:06 \\ ML 5.0\end{tabular}  & \begin{tabular}[c]{@{}c@{}}26-Jan-2012 09:57 \\ ML 2.9\end{tabular} & \begin{tabular}[c]{@{}c@{}}25-Jan-2012 08:06 \\ ML 5.0\end{tabular} &  \\ \hline
34                  & 339                & \begin{tabular}[c]{@{}c@{}}20-May-2012 02:03 \\ ML 5.9\end{tabular}  & \begin{tabular}[c]{@{}c@{}}14-Sep-2012 02:47 \\ ML 3.0\end{tabular} & \begin{tabular}[c]{@{}c@{}}20-May-2012 02:03 \\ ML 5.9\end{tabular} &  \\ \hline
37                  & 2                  & \begin{tabular}[c]{@{}c@{}}03-Oct-2012 14:41 \\ ML 4.5\end{tabular}  & \begin{tabular}[c]{@{}c@{}}03-Oct-2012 17:18 \\ ML 3.2\end{tabular} & \begin{tabular}[c]{@{}c@{}}03-Oct-2012 14:41 \\ ML 4.5\end{tabular} &  \\ \hline
38                  & 31                 & \begin{tabular}[c]{@{}c@{}}25-Oct-2012 23:05 \\ ML 5.0\end{tabular}  & \begin{tabular}[c]{@{}c@{}}18-Dec-2012 18:14 \\ ML 2.9\end{tabular} & \begin{tabular}[c]{@{}c@{}}25-Oct-2012 23:05 \\ ML 5.0\end{tabular} &  \\ \hline
41                  & 3                  & \begin{tabular}[c]{@{}c@{}}25-Jan-2013 14:48 \\ ML 4.8\end{tabular}  & \begin{tabular}[c]{@{}c@{}}20-Feb-2013 21:39 \\ ML 3.0\end{tabular} & \begin{tabular}[c]{@{}c@{}}25-Jan-2013 14:48 \\ ML 4.8\end{tabular} &  \\ \hline
42                  & 2                  & \begin{tabular}[c]{@{}c@{}}16-Feb-2013 21:16 \\ ML 4.7\end{tabular}  & \begin{tabular}[c]{@{}c@{}}23-Feb-2013 17:17 \\ ML 3.2\end{tabular} & \begin{tabular}[c]{@{}c@{}}16-Feb-2013 21:16 \\ ML 4.7\end{tabular} &  \\ \hline
43                  & 36                 & \begin{tabular}[c]{@{}c@{}}21-Jun-2013 10:33 \\ ML 5.2\end{tabular}  & \begin{tabular}[c]{@{}c@{}}13-Sep-2013 02:10 \\ ML 3.0\end{tabular} & \begin{tabular}[c]{@{}c@{}}21-Jun-2013 10:33 \\ ML 5.2\end{tabular} &  \\ \hline
44                  & 6                  & \begin{tabular}[c]{@{}c@{}}21-Jul-2013 01:32 \\ ML 4.9\end{tabular}  & \begin{tabular}[c]{@{}c@{}}04-Oct-2013 15:43 \\ ML 3.3\end{tabular} & \begin{tabular}[c]{@{}c@{}}21-Jul-2013 01:32 \\ ML 4.9\end{tabular} &  \\ \hline
47                  & 12                 & \begin{tabular}[c]{@{}c@{}}29-Dec-2013 17:08 \\ ML 5.0\end{tabular}  & \begin{tabular}[c]{@{}c@{}}20-Jan-2014 07:55 \\ ML 3.7\end{tabular} & \begin{tabular}[c]{@{}c@{}}29-Dec-2013 17:08 \\ ML 5.0\end{tabular} &  \\ \hline
50                  & 3                  & \begin{tabular}[c]{@{}c@{}}07-Apr-2014 19:26 \\ ML 4.7\end{tabular}  & \begin{tabular}[c]{@{}c@{}}14-Jul-2014 03:09 \\ ML 3.3\end{tabular} & \begin{tabular}[c]{@{}c@{}}07-Apr-2014 19:26 \\ ML 4.7\end{tabular} &  \\ \hline
58                  & 2                  & \begin{tabular}[c]{@{}c@{}}08-Feb-2016 15:35 \\ ML 4.6\end{tabular}  & \begin{tabular}[c]{@{}c@{}}08-Feb-2016 17:57 \\ ML 3.7\end{tabular} & \begin{tabular}[c]{@{}c@{}}08-Feb-2016 15:35 \\ ML 4.6\end{tabular} &  \\ \hline
60                  & 1346               & \begin{tabular}[c]{@{}c@{}}24-Aug-2016 01:36 \\ ML 6.0\end{tabular}  & \begin{tabular}[c]{@{}c@{}}04-May-2017 10:13 \\ ML 3.2\end{tabular} & \begin{tabular}[c]{@{}c@{}}30-Oct-2016 06:40 \\ ML 6.1\end{tabular} &  \\ \hline
65                  & 23                 & \begin{tabular}[c]{@{}c@{}}10-Apr-2018 03:11 \\ ML 4.7\end{tabular}  & \begin{tabular}[c]{@{}c@{}}05-Jul-2018 07:19 \\ ML 3.1\end{tabular} & \begin{tabular}[c]{@{}c@{}}10-Apr-2018 03:11 \\ ML 4.7\end{tabular} & \\ \hline
67                  & 14                 & \begin{tabular}[c]{@{}c@{}}14-Aug-2018 21:48\\ ML 4.7\end{tabular}   & \begin{tabular}[c]{@{}c@{}}25-Aug-2018 15:54 \\ ML 3.3\end{tabular} & \begin{tabular}[c]{@{}c@{}}16-Aug-2018 18:19 \\ ML 5.2\end{tabular} & \\  \hline
69                  & 39                 & \begin{tabular}[c]{@{}c@{}}06-Oct-2018 00:34 \\ ML 4.8\end{tabular}  & \begin{tabular}[c]{@{}c@{}}07-Jan-2019 01:31 \\ ML 2.9\end{tabular} & \begin{tabular}[c]{@{}c@{}}06-Oct-2018 00:34 \\ ML 4.8\end{tabular} &  \\ \hline
70                  & 2                  & \begin{tabular}[c]{@{}c@{}}14-Jan-2019  23:03 \\ ML 4.6\end{tabular} & \begin{tabular}[c]{@{}c@{}}14-Jan-2019 23:29 \\ ML 3.0\end{tabular} & \begin{tabular}[c]{@{}c@{}}14-Jan-2019 23:03 \\ ML 4.6\end{tabular} & \\  \hline
74                  & 12                 & \begin{tabular}[c]{@{}c@{}}09-Dec-2019 03:37 \\ ML 4.5\end{tabular}  & \begin{tabular}[c]{@{}c@{}}14-Dec-2019 16:55 \\ ML 3.1\end{tabular} & \begin{tabular}[c]{@{}c@{}}09-Dec-2019 03:37 \\ ML 4.5\end{tabular} & 
\end{tabular}
\end{table}

\begin{table}[]
\caption{GK-clusters with the strongest magnitude ML $\ge$ 4.5 (UTC time).\vspace{0.5cm}}
\label{tab:tableS3}
\hspace{-1cm}
\begin{tabular}{cccccl}
\textbf{Cluster ID} & \textbf{NR events} & \textbf{First event}                                                 & \textbf{Last event}                                                 & \textbf{Strongest event}                                             &  \\ \hline
14                  & 16                 & \begin{tabular}[c]{@{}c@{}}23-Dec-2008 15:24\\ ML 5.2\end{tabular}   & \begin{tabular}[c]{@{}c@{}}18-Jun-2009 11:55 \\ ML 3.6\end{tabular} & \begin{tabular}[c]{@{}c@{}}23-Dec-2008 15:24\\ ML 5.2\end{tabular}   &  \\ \hline
15                  & 352                & \begin{tabular}[c]{@{}c@{}}30-Mar-2009 13:38 \\ ML 4.1\end{tabular}  & \begin{tabular}[c]{@{}c@{}}26-May-2010 23:38\\  ML 3.0\end{tabular} & \begin{tabular}[c]{@{}c@{}}06-Apr-2009 01:32\\ ML 5.9\end{tabular}   &  \\ \hline
16                  & 2                  & \begin{tabular}[c]{@{}c@{}}05-Apr-2009 20:20 \\ ML 4.6\end{tabular}  & \begin{tabular}[c]{@{}c@{}}06-Apr-2009 03:33 \\ ML 3.3\end{tabular} & \begin{tabular}[c]{@{}c@{}}05-Apr-2009 20:20 \\ ML 4.6\end{tabular}  &  \\ \hline
17                  & 3                  & \begin{tabular}[c]{@{}c@{}}07-Sep-2009 21:26 \\ ML 4.5\end{tabular}  & \begin{tabular}[c]{@{}c@{}}29-Oct-2009 07:25 \\ ML 3.6\end{tabular} & \begin{tabular}[c]{@{}c@{}}07-Sep-2009 21:26 \\ ML 4.5\end{tabular}  &  \\ \hline
26                  & 2                  & \begin{tabular}[c]{@{}c@{}}17-Sep-2010 12:20 \\ ML 4.5\end{tabular}  & \begin{tabular}[c]{@{}c@{}}21-Sep-2010 07:02\\ ML 2.9\end{tabular}  & \begin{tabular}[c]{@{}c@{}}17-Sep-2010 12:20 \\ ML 4.5\end{tabular}  &  \\ \hline
30                  & 4                  & \begin{tabular}[c]{@{}c@{}}17-Jul-2011 18:30 \\ ML 4.8\end{tabular}  & \begin{tabular}[c]{@{}c@{}}27-Jul-2011 01:13 \\ ML 3.0\end{tabular} & \begin{tabular}[c]{@{}c@{}}17-Jul-2011 18:30 \\ ML 4.8\end{tabular}  &  \\ \hline
35                  & 163                & \begin{tabular}[c]{@{}c@{}}25-Jan-2012 08:06 \\ ML 5.0\end{tabular}  & \begin{tabular}[c]{@{}c@{}}13-Nov-2012 15:09 \\ ML 3.0\end{tabular} & \begin{tabular}[c]{@{}c@{}}29-May-2012 10:55 \\ ML 5.3\end{tabular}  &  \\ \hline
39                  & 184                & \begin{tabular}[c]{@{}c@{}}20-May-2012 02:03 \\ ML 5.9\end{tabular}  & \begin{tabular}[c]{@{}c@{}}30-May-2013 01:49 \\ ML 2.9\end{tabular} & \begin{tabular}[c]{@{}c@{}}20-May-2012 02:03 \\ ML 5.9\end{tabular}  &  \\ \hline
42                  & 2                  & \begin{tabular}[c]{@{}c@{}}03-Oct-2012 14:41 \\ ML 4.5\end{tabular}  & \begin{tabular}[c]{@{}c@{}}03-Oct-2012 17:18 \\ ML 3.2\end{tabular} & \begin{tabular}[c]{@{}c@{}}03-Oct-2012 14:41 \\ ML 4.5\end{tabular}  &  \\ \hline
43                  & 34                 & \begin{tabular}[c]{@{}c@{}}25-Oct-2012 23:05 \\ ML 5.0\end{tabular}  & \begin{tabular}[c]{@{}c@{}}17-Mar-2013 14:22 \\ ML 3.0\end{tabular} & \begin{tabular}[c]{@{}c@{}}25-Oct-2012 23:05 \\ ML 5.0\end{tabular}  &  \\ \hline
45                  & 3                  & \begin{tabular}[c]{@{}c@{}}25-Jan-2013 14:48 \\ ML 4.8\end{tabular}  & \begin{tabular}[c]{@{}c@{}}20-Feb-2013 21:39 \\ ML 3.0\end{tabular} & \begin{tabular}[c]{@{}c@{}}25-Jan-2013 14:48 \\ ML 4.8\end{tabular}  &  \\ \hline
47                  & 2                  & \begin{tabular}[c]{@{}c@{}}16-Feb-2013 21:16 \\ ML 4.7\end{tabular}  & \begin{tabular}[c]{@{}c@{}}23-Feb-2013 17:17 \\ ML 3.2\end{tabular} & \begin{tabular}[c]{@{}c@{}}16-Feb-2013 21:16 \\ ML 4.7\end{tabular}  &  \\ \hline
49                  & 37                 & \begin{tabular}[c]{@{}c@{}}21-Jun-2013 10:33 \\ ML 5.2\end{tabular}  & \begin{tabular}[c]{@{}c@{}}25-Oct-2013 09:17 \\ ML 3.1\end{tabular} & \begin{tabular}[c]{@{}c@{}}21-Jun-2013 10:33 \\ ML 5.2\end{tabular}  &  \\ \hline
50                  & 6                  & \begin{tabular}[c]{@{}c@{}}21-Jul-2013 01:32 \\ ML 4.9\end{tabular}  & \begin{tabular}[c]{@{}c@{}}04-Oct-2013 15:43 \\ ML 3.3\end{tabular} & \begin{tabular}[c]{@{}c@{}}21-Jul-2013 01:32 \\ ML 4.9\end{tabular}  &  \\ \hline
53                  & 12                 & \begin{tabular}[c]{@{}c@{}}29-Dec-2013 17:08 \\ ML 5.0\end{tabular}  & \begin{tabular}[c]{@{}c@{}}20-Jan-2014 07:55 \\ ML 3.7\end{tabular} & \begin{tabular}[c]{@{}c@{}}29-Dec-2013 17:08 \\ ML 5.0\end{tabular}  &  \\ \hline
56                  & 4                  & \begin{tabular}[c]{@{}c@{}}07-Apr-2014 19:26 \\ ML 4.7\end{tabular}  & \begin{tabular}[c]{@{}c@{}}14-Jul-2014 03:09 \\ ML 3.3\end{tabular} & \begin{tabular}[c]{@{}c@{}}07-Apr-2014 19:26 \\ ML 4.7\end{tabular}  &  \\ \hline
64                  & 2                  & \begin{tabular}[c]{@{}c@{}}08-Feb-2016 15:35 \\ ML 4.6\end{tabular}  & \begin{tabular}[c]{@{}c@{}}08-Feb-2016 17:57 \\ ML 3.7\end{tabular} & \begin{tabular}[c]{@{}c@{}}08-Feb-2016 15:35 \\ ML 4.6\end{tabular}  &  \\ \hline
66                  & 1453               & \begin{tabular}[c]{@{}c@{}}24-Aug-2016 01:36  \\ ML 6.0\end{tabular} & \begin{tabular}[c]{@{}c@{}}17-May-2018 3:57 \\ ML 3.0\end{tabular}  & \begin{tabular}[c]{@{}c@{}}30-Oct-2016 06:40  \\ ML 6.1\end{tabular} &  \\ \hline
70                  & 15                 & \begin{tabular}[c]{@{}c@{}}14-Aug-2018 21:48 \\ ML 4.7\end{tabular}  & \begin{tabular}[c]{@{}c@{}}13-Oct-2018 01:55 \\ ML 3.0\end{tabular} & \begin{tabular}[c]{@{}c@{}}16-Aug-2018 18:19 \\ ML 5.2\end{tabular}  &  \\ \hline
72                  & 45                 & \begin{tabular}[c]{@{}c@{}}06-Oct-2018 00:34 \\ ML 4.8\end{tabular}  & \begin{tabular}[c]{@{}c@{}}08-Jan-2019 23:50 \\ ML 4.1\end{tabular} & \begin{tabular}[c]{@{}c@{}}06-Oct-2018 00:34 \\ ML 4.8\end{tabular}  &  \\ \hline
73                  & 2                  & \begin{tabular}[c]{@{}c@{}}14-Jan-2019 23:03 \\ ML 4.6\end{tabular}  & \begin{tabular}[c]{@{}c@{}}14-Jan-2019 23:29 \\ ML 3.0\end{tabular} & \begin{tabular}[c]{@{}c@{}}14-Jan-2019 23:03 \\ ML 4.6\end{tabular}  &  \\ \hline
77                  & 12                 & \begin{tabular}[c]{@{}c@{}}09-Dec-2019 03:37 \\ ML 4.5\end{tabular}  & \begin{tabular}[c]{@{}c@{}}14-Dec-2019 16:55\\ ML 3.1\end{tabular}  & \begin{tabular}[c]{@{}c@{}}09-Dec-2019 03:37 \\ ML 4.5\end{tabular}  & 
\end{tabular}
\end{table}

\noindent\textbf{Table~\ref{tab:tableS2}.}
Some statistics, specified in the first column, of the events clustered in any ULG- or GK-cluster (ULG-IN, GK-IN), of the non-clustered events (ULG-OUT, GK-OUT), and of all the events together.

\begin{table}
\centering
\caption{Some statistics, specified in the first column, of the events clustered in any ULG- or GK-cluster (ULG-IN, GK-IN), of the non-clustered events (ULG-OUT, GK-OUT), and of all the events together. \vspace{0.5cm}}
\label{tab:tableS2}
\begin{tabular}{c|cc|cc|c}
\multicolumn{1}{l}{}                                                                 & \multicolumn{2}{c}{\textbf{ULG-clusters}} & \multicolumn{2}{c}{\textbf{GK- clusters}} & \textbf{ALL} \\
\multicolumn{1}{l}{}                                                                                  & \textbf{IN}         & \textbf{OUT}        & \textbf{IN}         & \textbf{OUT}        &                               \\ \hline
\textit{\begin{tabular}[c]{@{}c@{}}Number of\\ events\end{tabular}}                                   & 2516                & 2568                & 2653                & 2431                & 5084                          \\ \hline
\textit{\begin{tabular}[c]{@{}c@{}}Sum of\\ expected\\ descendants\end{tabular}}                      & 2497                & 1071                & 2542                & 1026                & 3568                          \\ \hline
\textit{\begin{tabular}[c]{@{}c@{}}Sum of Prob.\\ independence\end{tabular}}                          & 46.96               & 1459.74             & 80.21               & 1426.49             & 1506.7                        \\ \hline
\textit{\begin{tabular}[c]{@{}c@{}}Independence\\ Probability\\ $\le$ 0.1\end{tabular}}                   & 97.4\%              & 32.4\%              & 95.6\%              & 30.7\%              & 64.6\%           \\ \hline
\textit{\begin{tabular}[c]{@{}c@{}}Independence\\ Probability\\ 0.1 $<$ \&  $\le$ 0.9 \quad \end{tabular}} & 1.4\%               & 24\%                & 2.9\%               & 23.7\%              & 12.8\%                        \\ \hline
\textit{\begin{tabular}[c]{@{}c@{}}Independence\\ Probability\\ $>$ 0.9\end{tabular}}        & 1.2\%               & 43.6\%              & 1.5\%               & 45.6\%              & 22.6\%                       
\end{tabular}
\end{table}

\noindent\textbf{Tables~\ref{tab:tableS4} and~\ref{tab:tableS5}.}
Numerical results of the two checks for the ULG-clusters and the GK-clusters, respectively, with the strongest magnitude ML $\ge$ 4.5.

\begin{table}[]
\caption{Numerical results of the two checks for the ULG-clusters with the strongest magnitude ML $\ge$ 4.5.\vspace{0.5cm}}
\label{tab:tableS4}
\hspace{-1.25cm}
\begin{tabular}{c|c|c|c|c|c}
\textbf{ID cluster$\;$}                                                                                   & \textbf{\makecell{NR\\  Number evs.$\;$ \\ in cluster}}                                                             & \textbf{\makecell{S1\\  Sum expected$\;$ \\ descendants}}                                                          & \textbf{\makecell{TEST 1$\;$ \\ S1 / NR\\ (ok if $\sim$1)}}                                                                                      & \textbf{\makecell{S2\\ Sum Prob. \\  independence$\;$}}                                                                                                 & \textbf{\makecell{CHECK 2 \\ $|$S2 - 1$|$\\  (ok if $\sim$0)}}       

\\
\hline 
\makecell{9\\ 10\\ 11\\ 13\\ 25\\ 30\\ 34\\ 37\\ 38\\ 41\\ 42\\ 43\\ 44\\ 47\\ 50\\ 58\\ 60\\ 65\\ 67\\ 69\\ 70\\ 74} & \makecell{15\\ 340\\ 2\\ 2\\ 4\\ 3\\ 339\\ 2\\ 31\\ 3\\ 2\\ 36\\ 6\\ 12\\ 3\\ 2\\ 1346\\ 23\\ 14\\ 39\\ 2\\ 12} & \makecell{15\\ 361\\ 2\\ 3\\ 3\\ 3\\ 341\\ 1\\ 31\\ 2\\ 2\\ 38\\ 5\\ 11\\ 4\\ 1\\ 1350\\ 19\\ 14\\ 41\\ 1\\ 9} & \makecell{1\\ 1.06\\ 1\\ 1.50\\ 0.75\\ 1\\ 1.01\\ 0.50\\ 1\\ 0.67\\ 1\\ 1.06\\ 0.83\\ 0.92\\ 1.33\\ 0.50\\ 1\\ 0.83\\ 1\\ 1.05\\ 0.50\\ 0.75} & \makecell{0.06\\ 1.09\\ 0.99\\ 1\\ 0.01\\ 0.97\\ 0.06\\ 0.57\\ 0.23\\ 1.05\\ 0.99\\ 0.04\\ 0.02\\ 1\\ 0.28\\ 0\\ 6.57\\ 0.04\\ 0.02\\ 2.10\\ 1\\ 0} & \makecell{0.94\\ 0.09\\ 0.01\\ 0\\ 0.99\\ 0.03\\ 0.94\\ 0.43\\ 0.77\\ 0.05\\ 0.01\\ 0.96\\ 0.98\\ 0\\ 0.72\\ 1\\ 5.57\\ 0.96\\ 0.98\\ 1.10\\ 0\\ 1}
\end{tabular}
\end{table}

\begin{table}[]
\caption{Numerical results of the two checks for the GK-clusters with the strongest magnitude ML $\ge$ 4.5.\vspace{0.5cm}}
\label{tab:tableS5}
\hspace{-1.25cm}
\begin{tabular}{c|c|c|c|c|c}
\textbf{ID cluster$\;$}                                                                                   & \textbf{\makecell{NR\\  Number evs.$\;$ \\ in cluster}}                                                             & \textbf{\makecell{S1\\  Sum expected$\;$ \\ descendants}}                                                          & \textbf{\makecell{TEST 1$\;$ \\ S1 / NR\\ (ok if $\sim$1)}}                                                                                      & \textbf{\makecell{S2\\ Sum Prob. \\  independence$\;$}}                                                                                                 & \textbf{\makecell{CHECK 2 \\ $|$S2 - 1$|$\\  (ok if $\sim$0)}}       

\\
\hline 
\makecell{14\\ 15\\ 16\\ 17\\ 26\\ 30\\ 35\\ 39\\ 42\\ 43\\ 45\\ 47\\ 49\\ 50\\ 53\\ 56\\ 64\\ 66\\ 70\\ 72\\ 73\\ 77} & \makecell{16\\ 352\\ 2\\ 3\\ 2\\ 4\\ 163\\ 184\\ 2\\ 34\\ 3\\ 2\\ 37\\ 6\\ 12\\ 4\\ 2\\ 1453\\ 15\\ 45\\ 2\\ 12} & \makecell{15\\ 364\\ 2\\ 3\\ 1\\ 3\\ 117\\ 228\\ 1\\ 31\\ 2\\ 2\\ 38\\ 5\\ 11\\ 4\\ 1\\ 1405\\ 14\\ 44\\ 1\\ 9} & \makecell{0.94\\ 1.03\\ 1\\ 1\\ 0.50\\ 0.75\\ 0.72\\ 1.24\\ 0.50\\ 0.91\\ 0.67\\ 1\\ 1.03\\ 0.83\\ 0.92\\ 1\\ 0.50\\ 0.97\\ 0.93\\ 0.98\\ 0.50\\ 0.75} & \makecell{0.78\\ 3.20\\ 0.99\\ 1.93\\ 0.98\\ 0.01\\ 1.09\\ 0.67\\ 0.57\\ 1.83\\ 1.05\\ 0.99\\ 0.85\\ 0.02\\ 1\\ 1.21\\ 0\\ 13.61\\ 0.80\\ 2.71\\ 1\\ 0} & \makecell{0.22\\ 2.20\\ 0.01\\ 0.93\\ 0.02\\ 0.99\\ 0.09\\ 0.33\\ 0.43\\ 0.83\\ 0.05\\ 0.01\\ 0.15\\ 0.98\\ 0\\ 0.21\\ 1\\ 12.61\\ 0.20\\ 1.71\\ 0\\ 1}
\end{tabular}
\end{table}

\end{document}